\documentclass{article}
\usepackage{amsmath}
\usepackage{amssymb}
\usepackage{amsthm}
\usepackage{graphicx}
\usepackage{subfig}
\usepackage[affil-it]{authblk}
\usepackage{natbib}
\usepackage{bibentry}
\nobibliography*
\usepackage{hyperref}
\usepackage[top=1in, bottom=1in, left=1in, right=1in]{geometry}
\usepackage{appendix}
\usepackage{longtable}
\numberwithin{equation}{section}
\newtheorem{prop}{Property}

\author[1]{Wei Jiang}

\author[2]{Jing-Hao Xue}

\author[1]{Weichuan Yu%
  \thanks{Mail: \texttt{eeyu@ust.hk}}}
\affil[1]{Department of Electronic and Computer Engineering, The Hong Kong University of Science and Technology, Clear Water Bay, Kowloon, Hong Kong, China}
\affil[2]{Department of Statistical Science, University College London, London WC1E 6BT, U.K.}
\date{}
\title{Estimating Reproducibility in Genome-Wide Association Studies}
\begin{document}

\maketitle

\subsection*{\centering Abstract}
%[Abstract] 4-8 sentences. Where, what, why, how, results
{\em
Genome-wide association studies (GWAS) are widely used to discover genetic variants associated with diseases. To control false positives, all findings from GWAS need to be verified with additional evidences, even for associations discovered from a high power study. Replication study is a common verification method by using independent samples. An association is regarded as true positive with a high confidence when it can be identified in both primary study and replication study. Currently, there is no systematic study on the behavior of positives in the replication study when the positive results of primary study are considered as the prior information.

In this paper, two probabilistic measures named Reproducibility Rate ($RR$) and False Irreproducibility Rate ($FIR$) are proposed to quantitatively describe the behavior of primary positive associations (i.e. positive associations identified in the primary study) in the replication study. $RR$ is a conditional probability measuring how likely a primary positive association will also be positive in the replication study. This can be used to guide the design of replication study, and to check the consistency between the results of primary study and those of replication study. $FIR$, on the contrary, measures how likely a primary positive association may still be a true positive even when it is negative in the replication study. This can be used to generate a list of potentially true associations in the irreproducible findings for further scrutiny. The estimation methods of these two measures are given. Simulation results and real experiments show that our estimation methods have high accuracy and good prediction performance.
}

\section{Introduction}
%[Introduction-Background] Which situation the problem occurs? (Example); Introduce big problem in the field.
% background of GWAS
Genome-wide association studies (GWAS) were designed to detect genetic variations associated with diseases by genotyping single nucleotide polymorphisms (SNPs) in different individuals \citep{hirschhorn2005genome}. Compared to traditional candidate gene studies \citep{tabor2002candidate} which are based on pathway information, GWAS avoid the selection bias by genotyping a dense set of SNPs across the whole genome. Also, GWAS is more powerful than linkage analysis in detecting genetic variants contributing to disease risk with modest effect size \citep{risch1996future}.

% Current status: Multi-stage GWAS; Joint analysis(pooling strategy/meta analysis)
Since the first GWAS study on age-related macular degeneration (AMD) \citep{klein2005complement}, there have been about 2000 GWAS reports so far with 14609 associations showing genome wide significance ($p$-value $\leq 5\times 10^{-8}$) for 756 different diseases/traits (\citeauthor{HindorffWebresources}, accessed [2015.05.28]). Taking advantage of the pooling strategy and meta-analysis method, GWAS for common diseases are becoming more and more powerful \citep{evangelou2013meta,voight2010twelve,morris2012large,global2013discovery,locke2015genetic}.

%False positives, Extra experiments vs. Replication study
The basic analysis method used in GWAS is hypothesis testing \citep{balding2006tutorial}. In order to reduce false positives, extra evidences are essential to verify the discoveries. Commonly, there are two strategies used in GWAS to discover and examine associations: joint analysis and replication based analysis. Joint analysis uses all available GWAS data for the same disease in the same population to find associated SNPs, either by pooling multiple stage genotyping data or by combining test statistics with meta-analysis method. Afterwards, extra biological experiments are conducted to verify the associations. Replication based analysis splits the data into two parts, one for discovery (commonly called primary study) and the other for validation (commonly called replication study). With only a subset of available data being used in the primary study, replication based analysis is less powerful than joint analysis \citep{skol2006joint}. But it gives us an alternative way to examine findings without carrying out additional experiments \citep{chanock2007replicating, kraft2009replication}. Thus, replication based analysis is a common method of choice when facing budget or study design constraints.

In this paper, replication based analysis is our focus. For a reproduced association, suppose the type I error rates in the primary study and the replication study are $\alpha_1$ and $\alpha_2$, respectively, the probability of observing more extreme statistics is below $\alpha_1\alpha_2$ when the association doesn't exist. Since this is a very low probability, the reproduced association has a very high confidence to be true positive. For associations irreproduced in the replication study, we may suspect that they are false alarms, but usually we cannot say much more about them. However, given information from the primary study, people like to know more: how probable is a primary positive association (i.e. positive association identified in the primary study) to be confirmed in the replication study?  What's the probability that a primary positive association is still a true positive even if it fails to show significance in the replication study? To answer these two questions, we need a systematic study on the behavior of primary positives in the replication study. Unfortunately, there is no such a study as yet.

%[Introduction-Contributions] What's your innovation points solving the problem? (Bulleted list of contributions)
The aim of this paper is to systematically study primary positives in the replication study setting and to answer the two questions. Our contributions are listed as the following:
\begin{enumerate}
\item Reproducibility rate ($RR$) is proposed to quantify the probability that a primary positive association will be confirmed in the replication study. $RR$ can be used to guide the design of replication study.
\item False irreproducibility rate ($FIR$) is proposed to quantify the probability that a primary positive association is still a true association even when it cannot be confirmed in the replication study. $FIR$ can be used to discover potentially true associations in the irreproduced results.
\item Estimation methods are proposed for $RR$ and $FIR$ when the summary statistics of the primary study are available. In other words, $RR$ and $FIR$ can be estimated even before the replication study is carried out. This nice property allows us to explore all possibilities in the design of replication study.
\end{enumerate}

The rest of this paper is organized as follows. In section \ref{methods}, we will give the mathematical definitions of $RR$ and $FIR$ first. We will also derive the relationship among local false discovery rate ($fdr$, \citealt{efron2005local}), power, $RR$ and $FIR$. Then we will estimate $RR$ and $FIR$ using the Bayesian framework with a two-component mixture prior. In section \ref{results}, we will first show simulation results, which demonstrate that the estimation of $RR$ and $FIR$ works well when data agree with model assumptions. Then we will show the empirical results using the Type 2 Diabetes (T2D) data from DIAbetes Genetics Replication And Meta-analysis (DIAGRAM, \citealt{morris2012large}) and the Low Density Lipoprotein (LDL) cholesterol data from Global Lipids Genetics Consortium (GLGC, \citealt{global2013discovery}). We will also show other potential applications of $RR$ and $FIR$ in the same section. In section \ref{discussion}, we will discuss limitations of our current modeling and estimation method. These provide guidance for the future work. Section \ref{conclusion} concludes the paper.

\section{Method}\label{methods}
%[Methods] How to solve the problem with your idea in detail?
\subsection{$RR$ and $FIR$}
As illustration, we use $log(OR)$ test to identify associations. Here $log(OR)$ stands for logarithm of the odds ratio. The model can be easily generalized to quantitative trait with one-way fixed-effects ANOVA. Section \ref{results} gives an example of $RR$ and $FIR$ estimation for GWAS with quantitative trait.

In the replication based analysis of GWAS, let's assume study $j$ ($j=1, 2$ denote primary study and replication study, respectively) has $n^{(j)}$ individuals, where $n_0^{(j)}$ of them are controls and $n_1^{(j)}$ are cases. The number of SNPs is $m$. We use $\pi_0$ to denote the proportion of null SNPs, which have no association with the disease.

For each SNP, we use $A$ to represent the non-effect allele, and $a$ to denote the effect allele. Table \ref{contTable} shows a contingency table of alleles. Using the contingency table, we can estimate the logarithm of the odds ratio
\begin{equation}
\hat{\mu}^{(j)}=\log n_{00}^{(j)}-\log n_{01}^{(j)}-\log n_{10}^{(j)}+\log n_{11}^{(j)}.
\end{equation}
The true effect size $\mu$ is normally unknown. Using Woolf's method, we can approximate the asymptotic standard error of $\hat{\mu}^{(j)}$ (denoted as $\sigma^{(j)}$) as \citep{woolf1955estimating}:
\begin{equation}
\sigma^{(j)}\approx \sqrt{\frac{1}{n_{00}^{(j)}}+\frac{1}{n_{01}^{(j)}}+\frac{1}{n_{10}^{(j)}}+\frac{1}{n_{11}^{(j)}}}.
\end{equation}
The null and alternative hypotheses are
\begin{equation}
\mathcal{H}_0:\ \mu=0,\text{ vs. } \mathcal{H}_1:\ \mu\neq 0.
\end{equation}
The corresponding test statistic is $z^{(j)}=\hat{\mu}^{(j)}/\sigma^{(j)}$. Let's assume the significance levels in two studies are $\alpha_1$ and $\alpha_2$, respectively.
\begin{table}[h!]\centering
\begin{tabular}{c|cc|c}
& A & a & Total\\ \hline
Control & $n_{00}^{(j)}$ & $n_{01}^{(j)}$ & $2n_0^{(j)}$\\
Case & $n_{10}^{(j)}$ & $n_{11}^{(j)}$ & $2n_1^{(j)}$\\ \hline
Total & $n_{00}^{(j)}+n_{10}^{(j)}$ & $n_{01}^{(j)}+n_{11}^{(j)}$ & $2n^{(j)}$
\end{tabular}
\caption{Contingency table of one SNP in study $j$. Please see the main text for explanation of the notations.}\label{contTable}
\end{table}

Since two-sided test is used in the primary study, a SNP showing association with the disease has the absolute value of its $z$-value larger than $z_{\alpha_1/2}$, i.e. $|z^{(1)}|>z_{\alpha_1/2}$, where $z_{u}$ is the upper $u$ quantile of the standard normal distribution ($0\leq u \leq 0.5$). For the positive SNP confirmed in the replication study, its $z$-value should be consistent with the $z$-value in the primary study. Thus, $z^{(2)}$ should have the same sign as $z^{(1)}$, and should be also larger than $z_{\alpha_2}$ in terms of absolute value, i.e. $sgn(z^{(1)})z^{(2)}>z_{\alpha_2}$, where the sign function reads
\begin{equation}
sgn(x)=\left\{ \begin{array}{ll}
1 & \text{if $x>0$}\\
0 & \text{if $x=0$}\\
-1 & \text{if $x<0$}
\end{array} \right. .
\end{equation}
The reason that the critical value is $z_{\alpha_2}$ instead of $z_{\alpha_2/2}$ is that the sign of the rejection region is fixed, and the test can be regarded as one-sided.

For a SNP revealing association in the primary study, $RR$ is defined as
\begin{equation}
RR= P(sgn(z^{(1)})Z^{(2)}>z_{\alpha_2} \big| z^{(1)}), \text{ where $|z^{(1)}|>z_{\alpha_1/2}$}.
\end{equation}
Correspondingly, $FIR$ is defined as
\begin{equation}
FIR=P(\mathcal{H}_1 \big| sgn(z^{(1)})Z^{(2)}\leq z_{\alpha_2}, z^{(1)}), \text{ where $|z^{(1)}|>z_{\alpha_1/2}$}.
\end{equation}

Bayes formula can be used to derive the relationship among $RR$, $FIR$, local false discovery rate of the primary study $fdr^{(1)}$ and power function of the replication study $\beta^{(2)}(\mu)$ (Details are in the appendix):
\begin{equation}
\arraycolsep=1.4pt\def\arraystretch{2.2}
\begin{array}{ll}
RR=fdr^{(1)} \alpha_2+(1-fdr^{(1)}) \eta^{(2)}\\
FIR=\dfrac{(1-fdr^{(1)}) (1-\eta^{(2)})}{1-RR}
\end{array},\label{relation}
\end{equation}
where $fdr^{(1)}=P(\mathcal{H}_0\big| z^{(1)})$, and $\eta^{(2)}=P(sgn(z^{(1)})Z^{(2)}>z_{\alpha_2}\big| z^{(1)}, \mathcal{H}_1)=E(\beta^{(2)}(\mu)\big| z^{(1)}, \mathcal{H}_1)$ is the Bayesian predictive power \citep{lecoutre2001bayesian} of the replication study, which averages the power among all possible effect size values given the test statistic in the primary study.

As indicated by Eq. (\ref{relation}), $RR$ can be regarded as a weighted average between true null component $\alpha_2$ and true associated component $\eta^{(2)}$, where $fdr^{(1)}$ and $1-fdr^{(1)}$ are the weights, respectively. $FIR$ is the proportion of weighted true associated component $(1-fdr^{(1)}) (1-\eta^{(2)})$ in the irreproducibility rate, namely
\begin{equation}
1-RR=fdr^{(1)} (1-\alpha_2)+(1-fdr^{(1)}) (1-\eta^{(2)}).
\end{equation}
Thus, the calculation of $RR$ and $FIR$ can be done once $fdr^{(1)}$ and $\eta^{(2)}$ are known.

Both $fdr^{(1)}$ and $\eta^{(2)}$ are the posterior probabilities which depend on the distribution of underlying true effect size value $\mu$. We need to specify a prior distribution of $\mu$ for the calculation of $fdr^{(1)}$ and $\eta^{(2)}$. In the following subsection, we will use a two-component mixture prior for $\mu$ to derive their calculation formulas.

\subsection{Two-component mixture prior}
In each study, the $log(OR)$ estimator $\hat{\mu}^{(j)}$ can be assumed normally distributed with a mean $\mu$ and a standard deviation $\sigma^{(j)}$, i.e.
\begin{eqnarray}
\frac{\hat{\mu}^{(j)}-\mu}{\sigma^{(j)}}\sim N(0,1).
\end{eqnarray}

The true effect size $\mu$ is unknown. Researches on heritability decomposition \citep{yang2010common} and effect size distribution \citep{park2010estimation} suggest that SNPs with small effect sizes occupy a larger proportion in the associated SNPs than those with large effect sizes. Hence, a natural prior for the effect size of the associated SNPs is a Gaussian prior with mean zero. Since we don't know whether an arbitrary SNP is associated or not, the following two-component mixture prior is used for all SNPs:
\begin{equation}
\mu\sim \pi_0 \delta_0+(1-\pi_0) N(0, \sigma_0^2),\label{prior}
\end{equation}
where $\delta_0$ is the distribution with point mass on zero whose probability density function (pdf) is Dirac function $\delta(x)$.

The local false discovery rate of the primary study can be calculated as:
\begin{equation}
fdr^{(1)}= \frac{\pi_0 \phi(z^{(1)}) }{\pi_0 \phi(z^{(1)}) +\pi_1 \phi ( \frac{z^{(1)} }{\sqrt{1+(\sigma_0/\sigma^{(1)})^2}} )},
\end{equation}
where $\phi(x)$ is the pdf of the standard normal distribution.

The Bayesian predictive power of the replication study can be calculated as follows (Details in the appendix):
\begin{equation}
\eta^{(2)}= \Phi(\frac{sgn(z^{(1)})z^*-z_{\alpha_2}}{\sigma^*}),
\end{equation}
where $z^*=\lambda \frac{\widehat{\mu}^{(1)}}{\sigma^{(2)}}$, $\sigma^*=\sqrt{1+\lambda\left(\frac{\sigma^{(1)}}{\sigma^{(2)}}\right)^2}$, $\lambda=\frac{1}{1+(\sigma^{(1)}/\sigma_0)^2}$ and $\Phi(x)$ is the cumulative density function (cdf) of the standard normal distribution.

When summary statistics of the primary study are available, the asymptotic standard error of $\hat{\mu}^{(2)}$ can be approximated by substituting observed allele frequencies from the
primary study into Woolf's method:
\begin{equation}
\sigma^{(2)}\approx \sqrt{\frac{n_0^{(1)}}{n_0^{(2)}}\left(\frac{1}{n_{00}^{(1)}}+\frac{1}{n_{01}^{(1)}}\right)+\frac{n_1^{(1)}}{n_1^{(2)}}\left(\frac{1}{n_{10}^{(1)}}+\frac{1}{n_{11}^{(1)}}\right)}.
\end{equation}

\subsection{Estimation}
Clearly, $RR$, $FIR$, $fdr^{(1)}$ and $\eta^{(2)}$ depend on hyperparameters $\pi_0$ and $\sigma_0$. Since all SNPs are assumed to share the same structure of distribution in terms of effect size in Eq. (\ref{prior}), the hyperparameters can be estimated with the test statistics of the primary study.

The estimation of $\pi_0$ has been addressed in the literature of $FDR$ control from the Bayesian point of view \citep{storey2003statistical}. Suppose there is a ``zero assumption'' that all SNPs with $p$-value$>\gamma$ have almost no chance to be truly associated SNPs. Let's denote the number of those SNPs as $m_+(\gamma)$. Then its expectation is
\begin{equation}
E(m_+(\gamma))=mP(p\text{-value}>\gamma, \mathcal{H}_0)=mP(\mathcal{H}_0)P(p\text{-value}>\gamma|\mathcal{H}_0)=m\pi_0(1-\gamma),
\end{equation}
which introduces an $\pi_0$ estimator
\begin{eqnarray}
\widehat{\pi}_0=\frac{m_{+}(\gamma)}{m(1-\gamma)}.
\end{eqnarray}
There is a tradeoff between bias and variance when choosing $\gamma$ in the estimation of $\pi_0$. \citeauthor{storey2003statistical} proposed a procedure without tuning the parameter $\gamma$. The automated procedure will evaluate $\hat{\pi}_0$ at different $\gamma$. Then, a natural cubic spline will fit to those evaluated values. The final $\hat{\pi}_0$ will be obtained at $\gamma=1$ of the fitted spline.

The estimator of $\sigma_0$ reads (see appendix for detail):
\begin{equation}
\widehat{\sigma}_0^2=\left(\frac{\sum_{i=1}^m (z^{(1)}_i)^2-m\pi_0}{(1-\pi_0)}-m\right)/\sum_{i=1}^m (1/\sigma^{(1)}_i)^2.
\end{equation}

For each SNP showing association in the primary study, Bootstrap can be used to obtain the confidence interval of $RR$ and $FIR$.

\section{Result}\label{results}
\subsection{Simulation experiments}
%[Results-Experiments] How to verify the method works?
% Description of simulation study
We use simulation experiments to check the following questions:
\begin{enumerate}
\item Can $RR$ and $FIR$ be accurately estimated?
\item How is the prediction performance of $\widehat{RR}$?
\begin{enumerate}
\item Can $\widehat{RR}$ predict whether a primary association will be reproduced in the replication study?
%\item How good is the prediction?
\item Can $\widehat{RR}$ describe the reproducibility probability well?
%\item Can $\widehat{RR}$ describe the probability of being reproduced for an association well?
\end{enumerate}
\item Can $\widehat{FIR}$ predict whether an irreproduced primary association is true positive or not?
\end{enumerate}

We simulate 2000 controls and 2000 cases in the primary study, and 1000 controls and 1000 cases in the replication study. The number of SNPs is $1\times 10^4$. The effect sizes of all SNPs are generated from the following two-component distribution:
\begin{equation}
\mu\sim 0.9\delta_0+0.1N(0,0.04).
\end{equation}
The minor allele frequencies are randomly simulated from a uniform distribution $U(0.05,0.5)$, and the prevalence of the disease is set to $1\%$. We use $\alpha_1=5\times 10^{-5}$ and $\alpha_2=5\times 10^{-3}$ as significance levels in the primary study and replication study, respectively.

\begin{figure}[h!]
    \centering
    \subfloat[$RR$]{\includegraphics[width=0.47\textwidth]{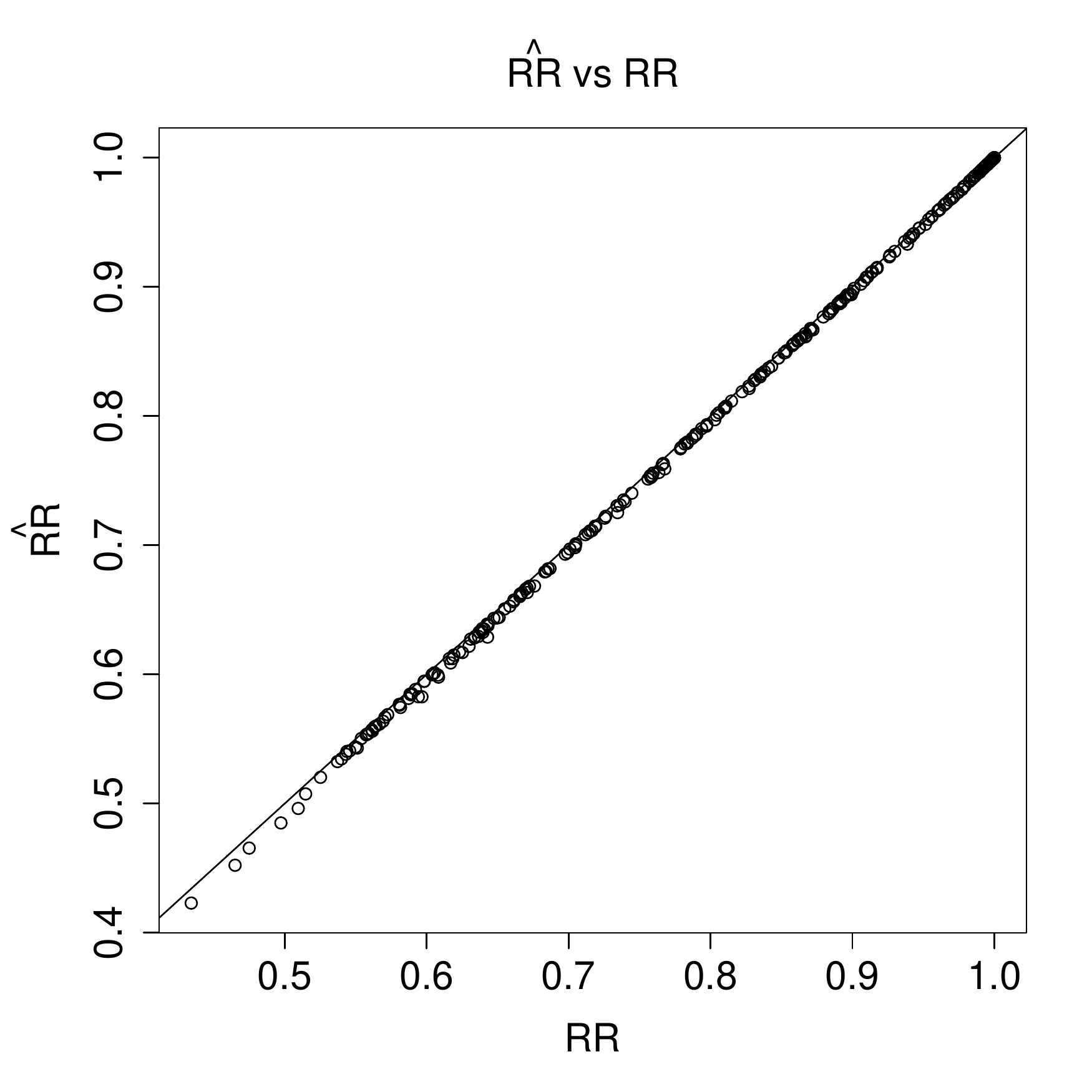}}
    \subfloat[$FIR$]{\includegraphics[width=0.47\textwidth]{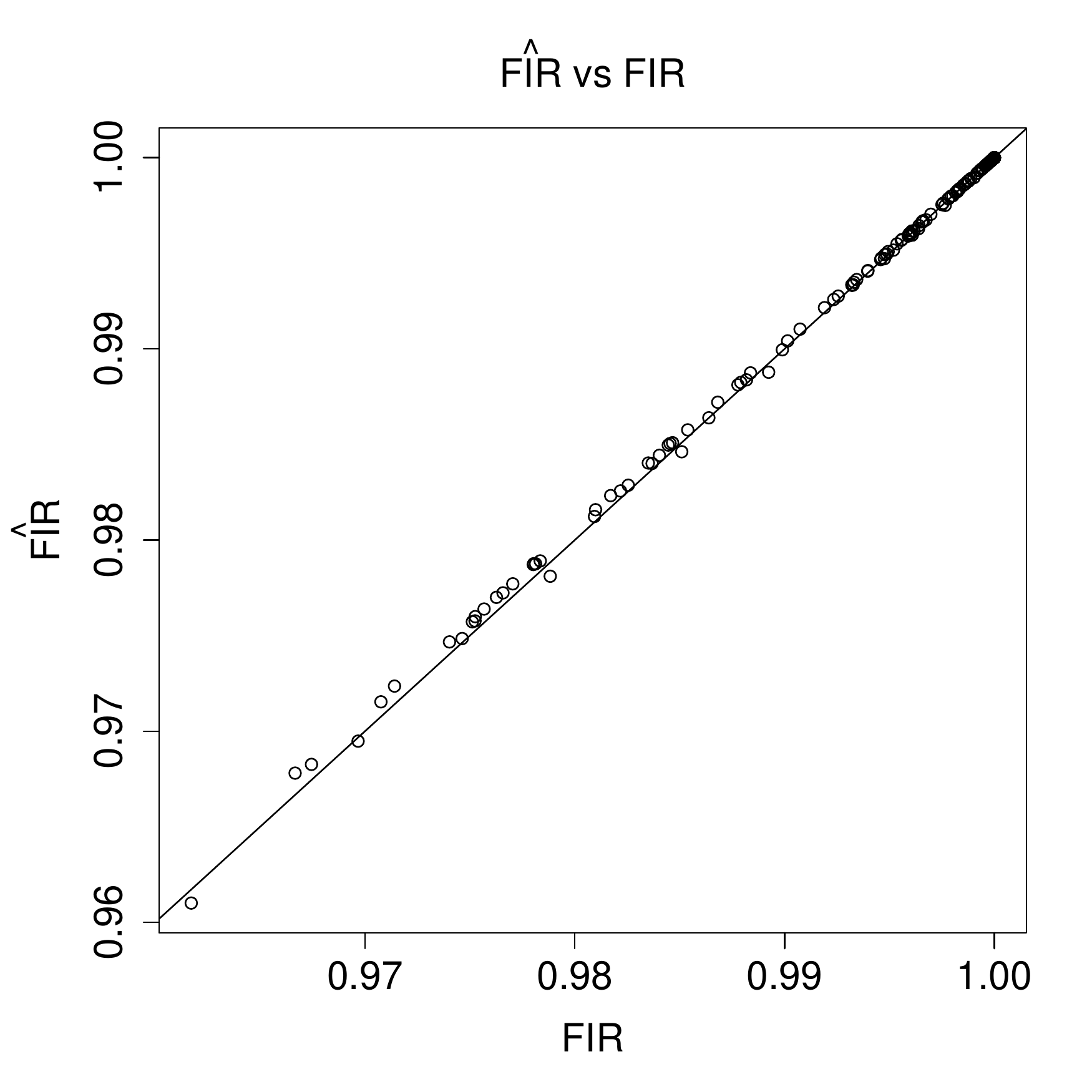}}
    \caption{$\widehat{RR}$ and $\widehat{FIR}$ can estimate $RR$ and $FIR$ accurately. The x-axis is the true values of $RR$ (in (a)) or $FIR$ (in (b)) in the simulation study, and the y-axis is the corresponding estimated values $\widehat{RR}$ (in (a)) or $\widehat{FIR}$ (in (b)). The solid line is $y=x$.}\label{estimation}
\end{figure}

%[Results] Does the method works? (details, data)
Figure \ref{estimation} shows the comparison between $\widehat{RR}$, $\widehat{FIR}$ and their true values. The two scatter plots show that both $\widehat{RR}$ and $\widehat{FIR}$ work well in terms of estimation accuracy. This kind of experiment has been run 5 times. The root mean square error of these two estimators in Table \ref{RMSE} show that $\widehat{RR}$ and $\widehat{FIR}$ have high estimation accuracy.

\begin{table}[h!]
\centering
\begin{tabular}{ccc}
  & RR & FIR \\
  \hline
run 1 & 0.004 & 0.000 \\
  run 2 & 0.017 & 0.002 \\
  run 3 & 0.000 & 0.001 \\
  run 4 & 0.010 & 0.001 \\
  run 5 & 0.013 & 0.004 \\
   \hline
Average & 0.009 & 0.002 \\
  \end{tabular}
\caption{Root mean square error of $\widehat{RR}$ and $\widehat{FIR}$ in the simulation experiments.}\label{RMSE}
\end{table}

\begin{figure}[h!]
    \centering
    \subfloat[PR curve of $\widehat{RR}$.]{\label{PRC_RR_sim} \includegraphics[width=0.47\textwidth]{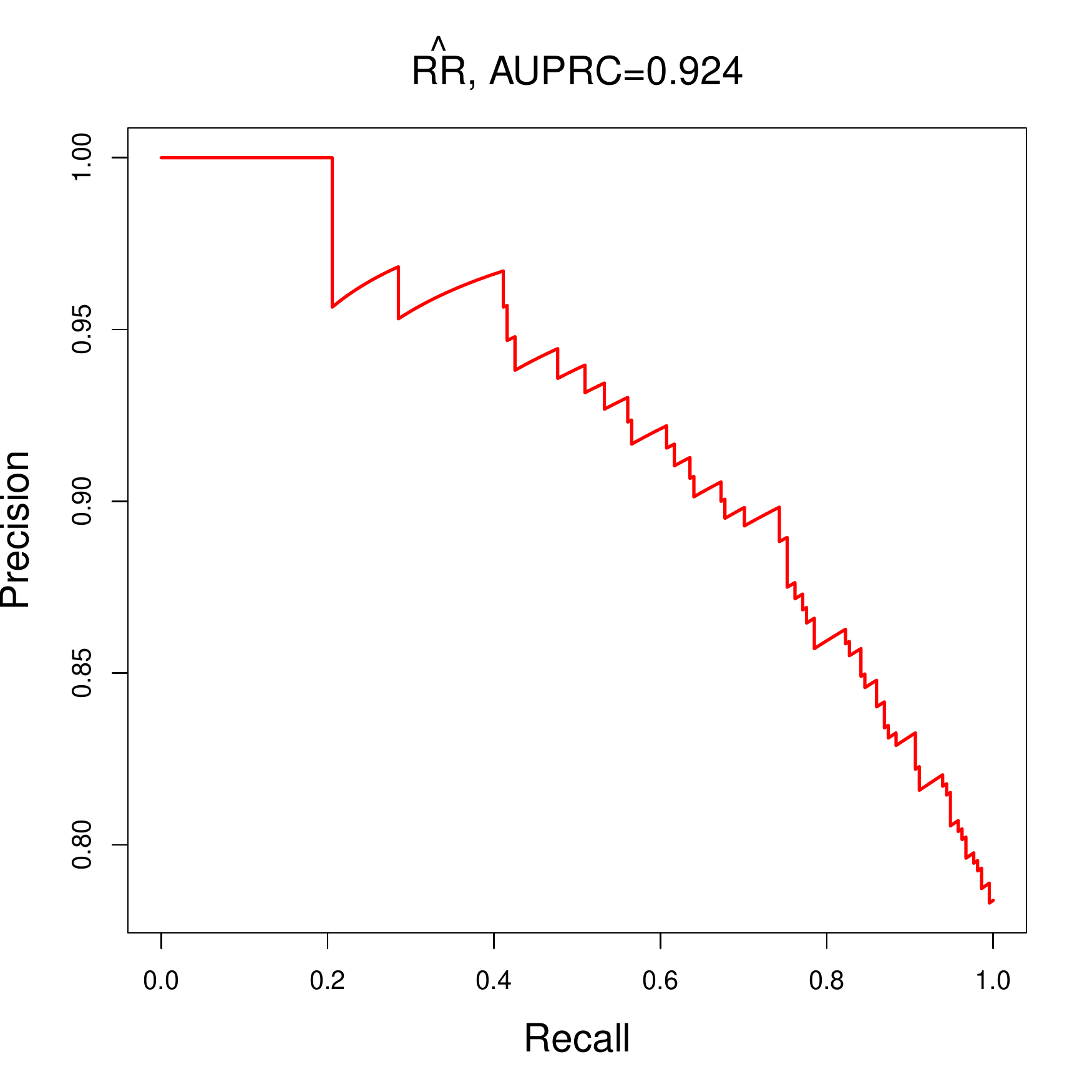}}
    \subfloat[Reproducibility Proportion (RP) vs. RR.]{\label{RP_RR_sim} \includegraphics[width=0.47\textwidth]{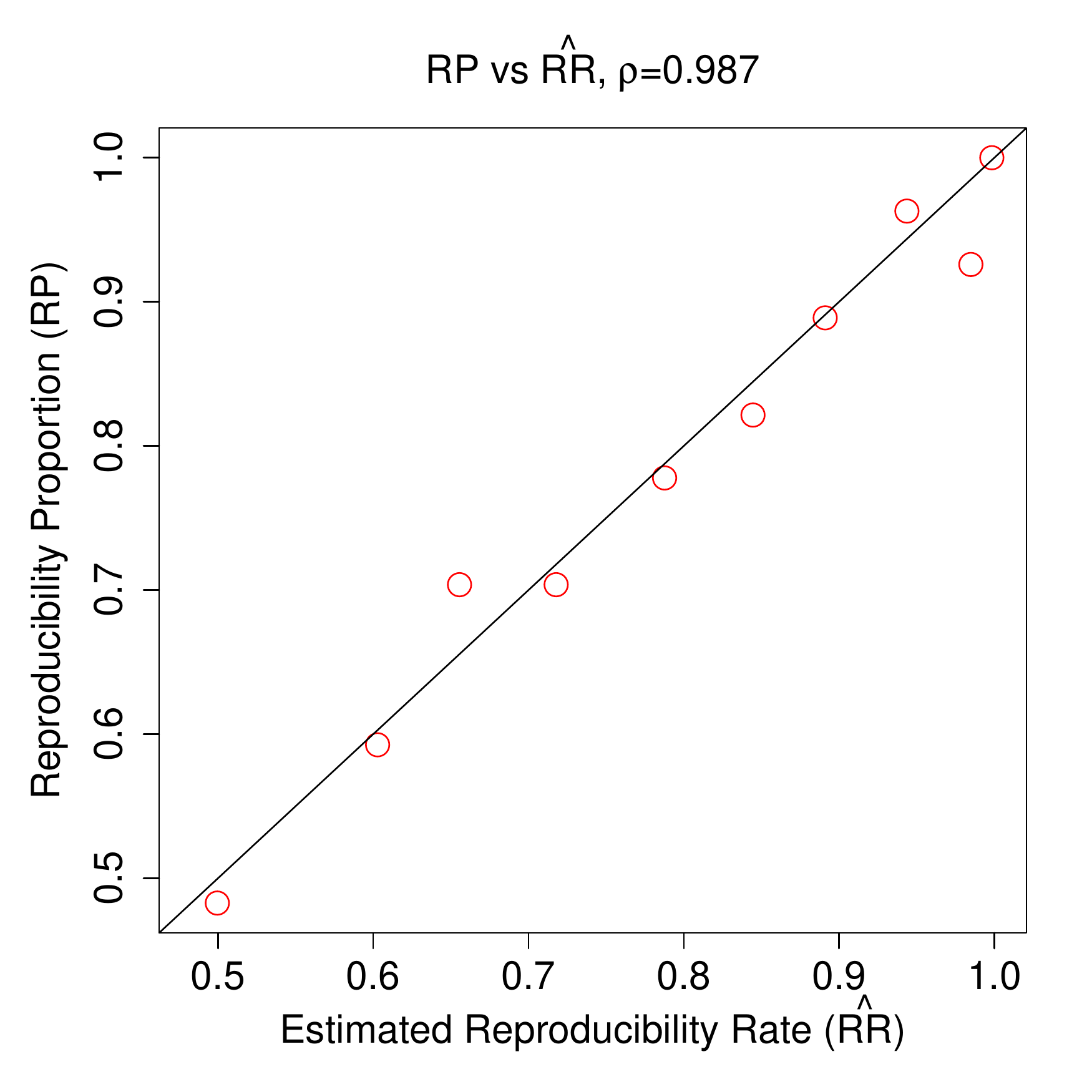}}
    \caption{$\widehat{RR}$ of an association can predict its reproducibility in the simulation study. (a) We use $\widehat{RR}$ as a score to decide reproduced/irreproduced status in the replication study. A PR curve is drawn by using different thresholds. The x-axis is the recall in reproducibility prediction in terms of $\widehat{RR}$, and the y-axis is the corresponding precision. $AUPRC$ is the area under precision-recall curve. (b) The associations are partitioned into 10 groups according to $\widehat{RR}$. The x-axis is the $\widehat{RR}$ of the group, which is the mid-point of the range of $\widehat{RR}$ within the group. The y-axis is the corresponding $RP$ of the group, which is the proportion of the reproduced associations in each group. The solid line is $y=x$.}
\end{figure}

In order to see whether $\widehat{RR}$ can predict the replication status well, we use $\widehat{RR}$ as a score to decide whether the association can be reproduced or not in the replication study. A Precision-Recall (PR) curve is drawn (Figure \ref{PRC_RR_sim}) by using different thresholds in the prediction. The reason that we choose PR curve instead of the commonly used Receiver Operator Characteristic (ROC) curve is that the replication status in the positive findings can be highly imbalanced, and PR curve can give more information about the prediction performance in the imbalanced situation \citep{davis2006relationship}. A high $\widehat{RR}$ value predicts that the association can be reproduced. The area under the PR curve is 0.924 in this simulation. This large area indicates that $\widehat{RR}$ has good prediction performance as an index about reproducibility.

We use the following procedure to see whether $\widehat{RR}$ can describe the reproducibility probability.
\begin{enumerate}
\item The associations are partitioned into groups according to $\widehat{RR}$. Each group has associations with approximately equal size. With 10 groups, the first group refers to $1/10$ of the associations having the highest $\widehat{RR}$, the second group refers to the next $1/10$ of the associations having the second decile of $\widehat{RR}$, and so on.
\item The proportion of the reproduced associations in each group is defined as reproducibility proportion ($RP$). The mid-point of the range of $\widehat{RR}$ is regarded as the $\widehat{RR}$ in this group. $RP$ and $\widehat{RR}$ are compared in each group.
\end{enumerate}
Figure \ref{RP_RR_sim} shows the comparison between $RP$ and $\widehat{RR}$ for 10 groups. We can see that these two quantities agree well. The correlation between them is 0.987. This result implies $\widehat{RR}$ can predict the reproducibility of findings.

\begin{figure}[h!]
    \centering
    \includegraphics[width=0.50\textwidth]{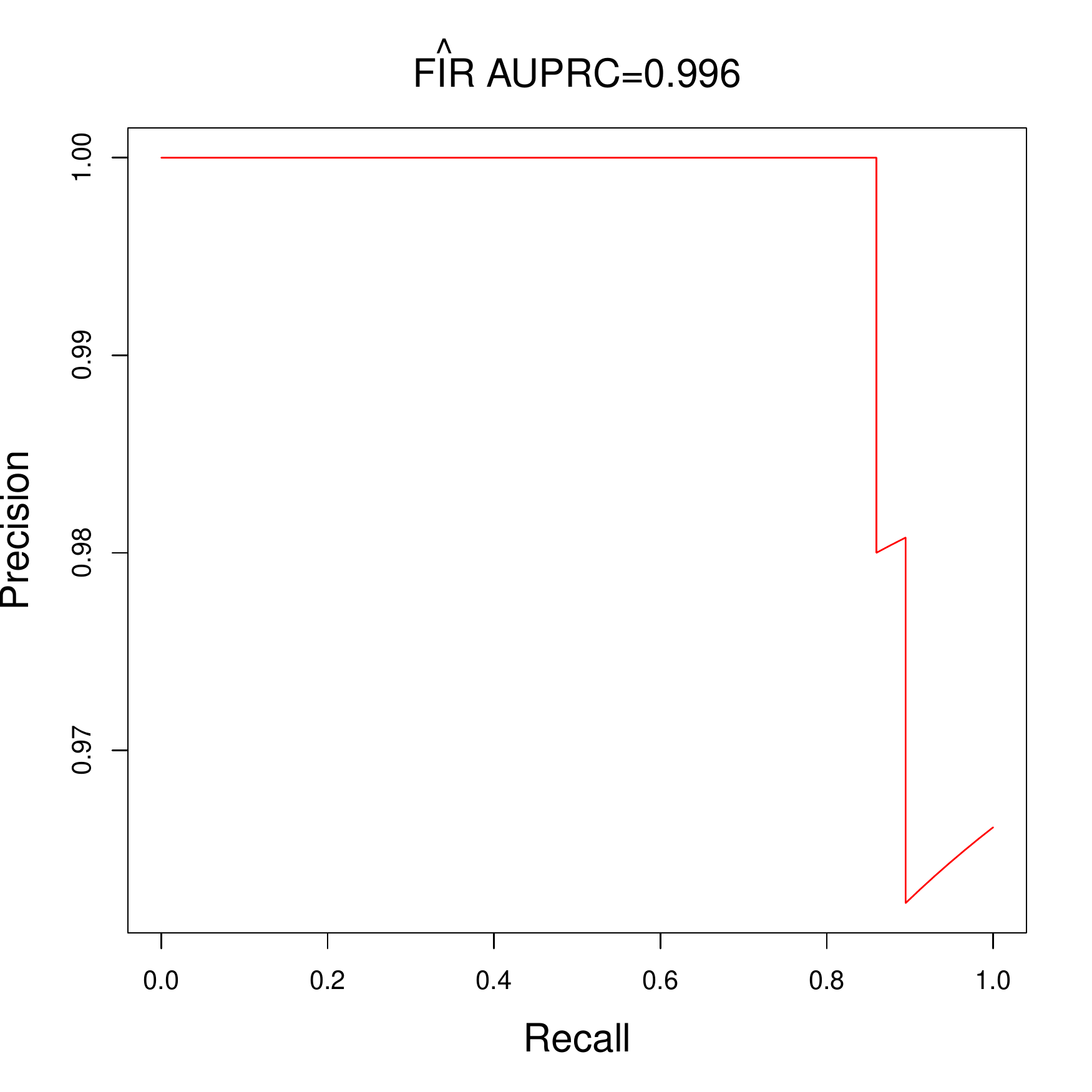}
    \caption{Precision-recall curve of $\widehat{FIR}$ in the simulation study. $\widehat{FIR}$ of an irreproduced finding can be a quantitative index to describe the potential that this finding is a true association. The x-axis is the recall in false irreproducibility prediction in terms of $\widehat{FIR}$, and the y-axis is the corresponding precision. $AUPRC$ is the area under precision-recall curve.}\label{PRC_FIR_sim}
\end{figure}

For the irreproduced findings, since we know whether it is true association or not in the simulation, a PR curve can be drawn for $\widehat{FIR}$ (Figure \ref{PRC_FIR_sim}). A high $\widehat{FIR}$ value predicts the irreproduced association to be a true association. The area under the cure is 0.998 in this simulation, which indicates $\widehat{FIR}$ has good prediction performance as an index about the potential being true association.

\subsection{T2D data from DIAGRAM}\label{T2D}
Public T2D dataset from DIAGRAM is used to further check our $RR$ estimation accuracy. For primary study, 56862 individuals are in the control group and 12171 individuals are in the case group. For replication study, the sample size in the control group is 55647, and the sample size in the case group is 21491. After filtering out SNPs with $p$-value$<0.01$ in the test of homogeneity, there are $m=89659$ SNPs remaining. The significance level used in the primary study is genome-wide significance level $5\times 10^{-8}$. The significance level used in the replication study is $5\times 10^{-6}$. The estimated proportion of null hypotheses is $\widehat{\pi}_0=0.924$, and the estimated effect size variance is $\widehat{\sigma}_0^2=5.43\times 10^{-3}$. There are $177$ SNPs showing significant associations in the primary study, from which 24 clumps are formed. Each clump contains SNPs in a nearby region ($<250kb$) with strong linkage disequilibrium between them ($r^2>0.5$). The SNP showing strongest association is selected to estimate $RR$ and $FIR$ in each clump. The results from all clumps are shown in the appendix.

\begin{figure}[h!]
    \centering
    \subfloat[Precision-recall curve.]{\label{PRC_RR_T2D}\includegraphics[width=0.47\textwidth]{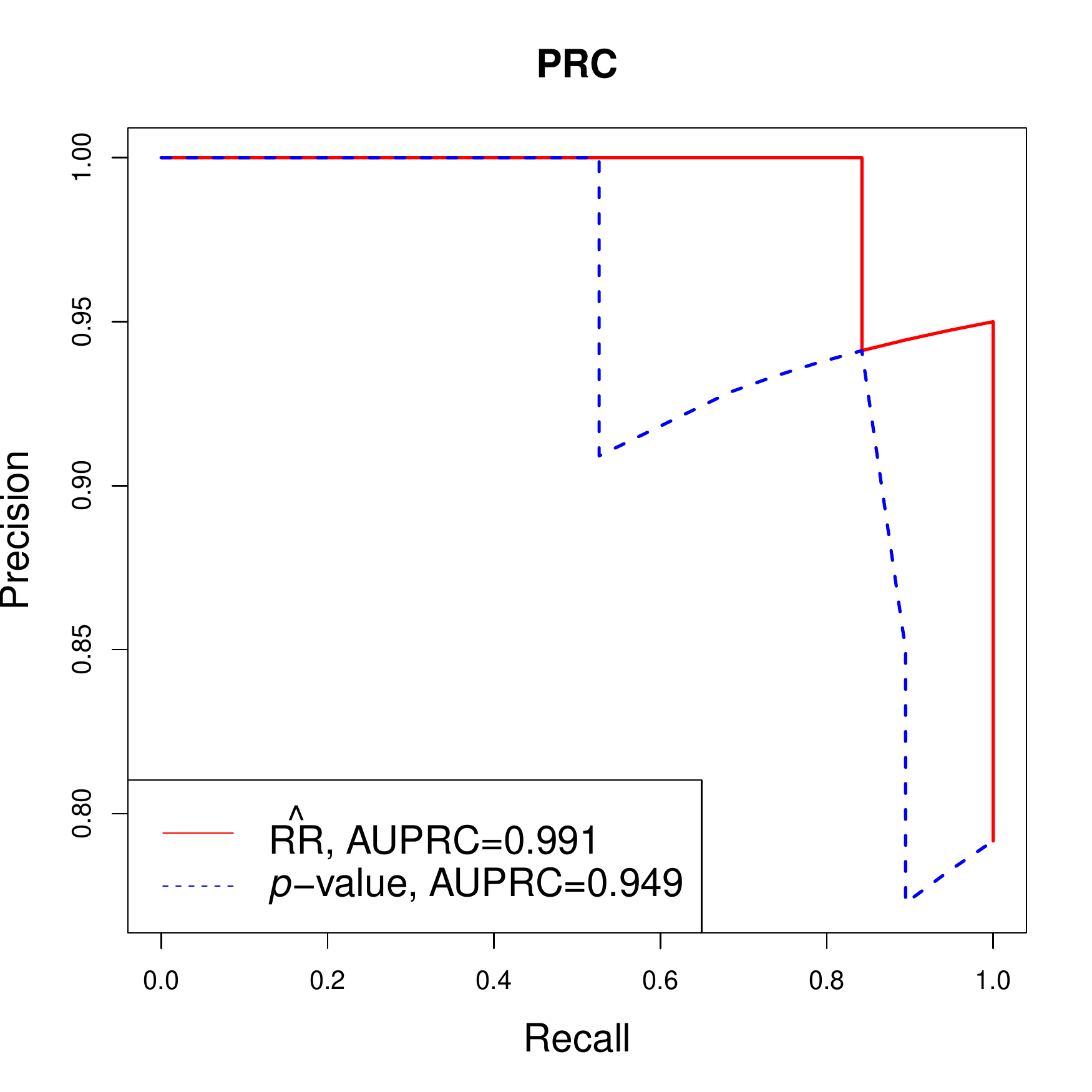}}
    \subfloat[Reproducibility Proportion (RP) vs. RR.]{\label{RP_RR_T2D}\includegraphics[width=0.47\textwidth]{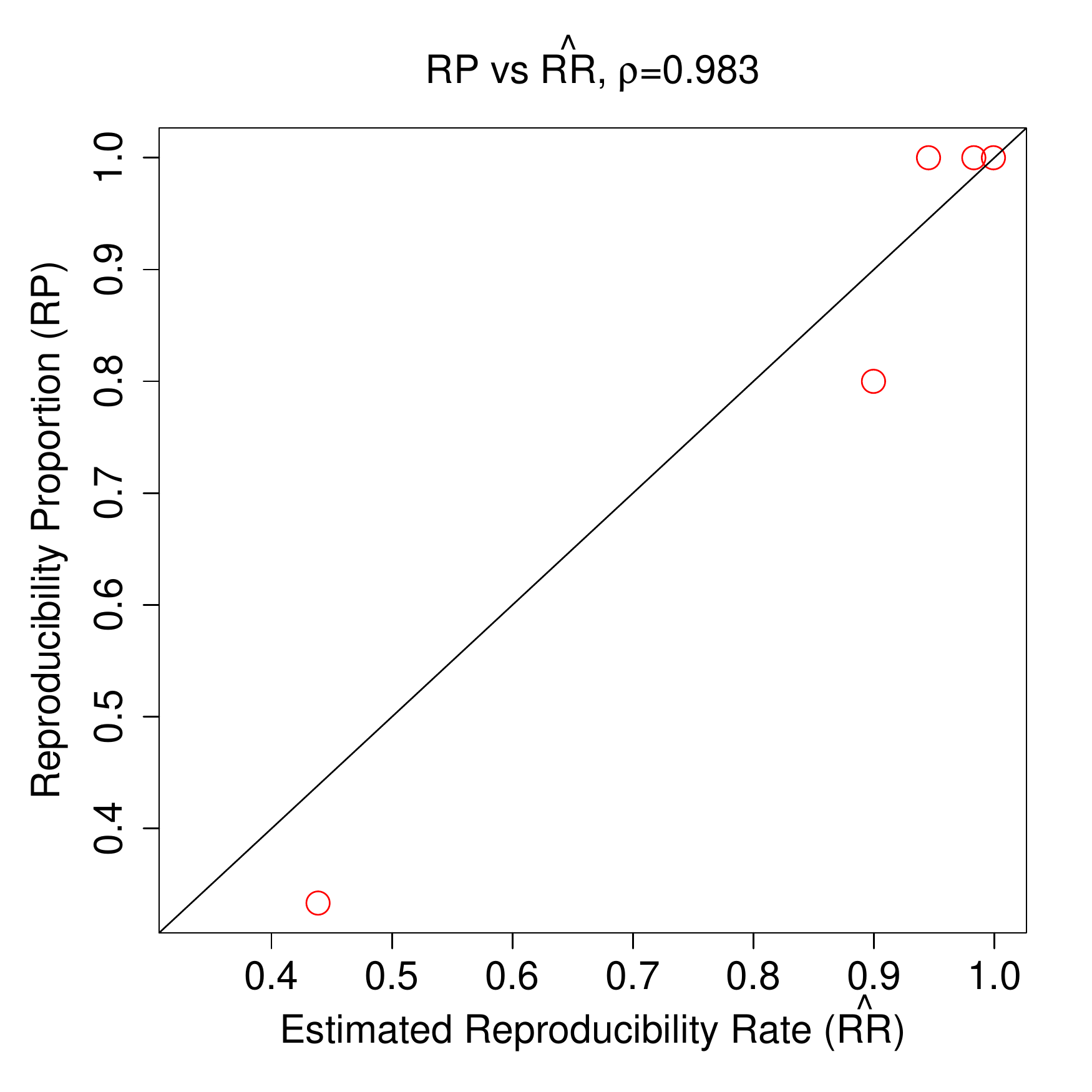}}
    \caption{Reproducibility prediction in T2D data from DIAGRAM. (a) The x-axis is the recall in reproducibility prediction in terms of $\widehat{RR}$, and the y-axis is the corresponding precision. $AUPRC$ is the area under precision-recall curve. Both PR curve based on $\widehat{RR}$ (solid line) and PR curve based on $p$-value (dashed line) are drawn in the figure. According to their $AUPRC$ values, $\widehat{RR}$ predicts reproducibility better than $p$-value. (b) The associations are partitioned into 5 groups according to $\widehat{RR}$. The x-axis is the $\widehat{RR}$ of the group, which is the mid-poin of $\widehat{RR}$ values. The y-axis is the corresponding $RP$ of the group, which is the proportion of the reproduced associations in each group. The solid line is $y=x$.}
\end{figure}

A precision-recall curve is drawn for the prediction of reproducibility based on $\widehat{RR}$ (Figure \ref{PRC_RR_T2D}). The area under the precision-recall curve is 0.991. In comparison, if $p$-value is regarded as an index describing reproducibility, then the area under the curve is 0.949, smaller than the area of $\widehat{RR}$. In order to see whether $\widehat{RR}$ can describe the reproducibility probability for an association, we calculate $RP$ by partitioning the clumps into 5 groups according to their $\widehat{RR}$ values, and make a comparison between $RP$ and $\widehat{RR}$. We use 5 groups here instead of 10 groups used in the simulation study because the number of the clumps is much smaller than the number of the associations in the simulation experiments. Figure \ref{RP_RR_T2D} shows the comparison between $RP$ and $\widehat{RR}$. These two quantities agree well. The correlation between them is 0.983. This result illustrates that $\widehat{RR}$ can well represent the probability of being reproduced for each of the findings in these data.

There are also $5$ irreproduced clumps in the result. While we may suspect that they are false positives if we follow the traditional strategy, their $\widehat{FIR}$ values indicate that they are all very likely to be true associations. To verify if this statement is true, we have used meta-analysis method to increase the power of the study. The corresponding $p$-values of these clumps are indeed smaller than the genome-wide significance level.

\subsection{LDL Cholesterol data from GLGC}\label{LDL}
We also conducted experiments in the published LDL cholesterol data from GLGC. The phenotype value measured in the study is quantitative. The estimated and standardized regression coefficients are used as test statistics. For primary study, there are about 93982 individuals. For replication study, the sample size is around 94565. After filtering out SNPs with $p$-value$<0.01$ in the test of homogeneity, there are $m=81942$ SNPs. The significance level in the primary study is $5\times 10^{-8}$, and the significance level in the replication study is $5\times 10^{-6}$. The estimated proportion of null hypotheses is $\widehat{\pi}_0=0.905$, and the estimated effect size variance is $\widehat{\sigma}_0^2=6.20\times 10^{-4}$. There are $748$ SNPs showing significant associations in the primary study, forming 161 clumps. The SNP showing strongest association is selected for estimating $RR$ and $FIR$ in each clump. The estimated results for all clumps are shown in the appendix.

\begin{figure}[h!]
    \centering
    \subfloat[Precision-recall curve.]{\label{PRC_RR_LDL}\includegraphics[width=0.47\textwidth]{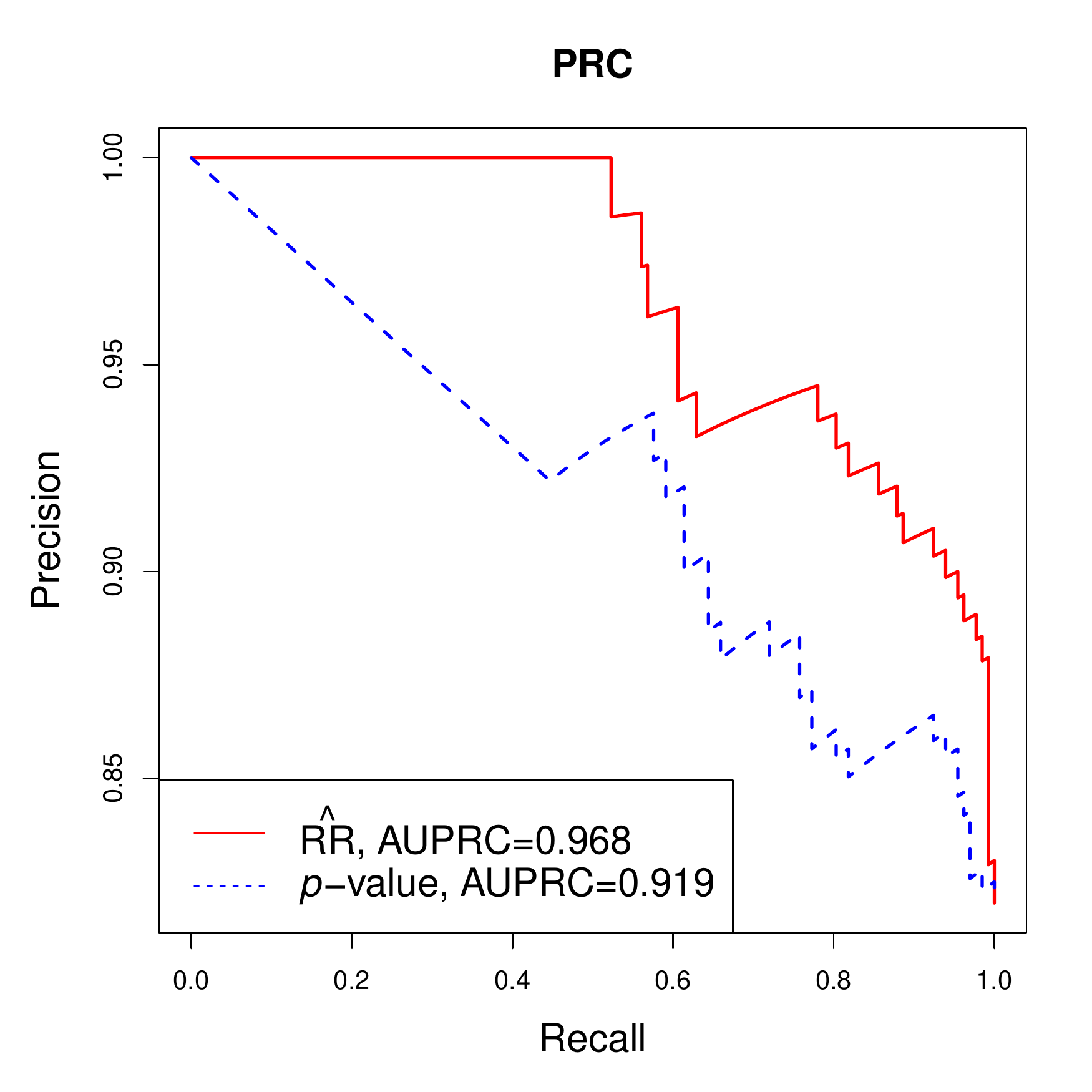}}
    \subfloat[Reproducibility Proportion (RP) vs. RR.]{\label{RP_RR_LDL}\includegraphics[width=0.47\textwidth]{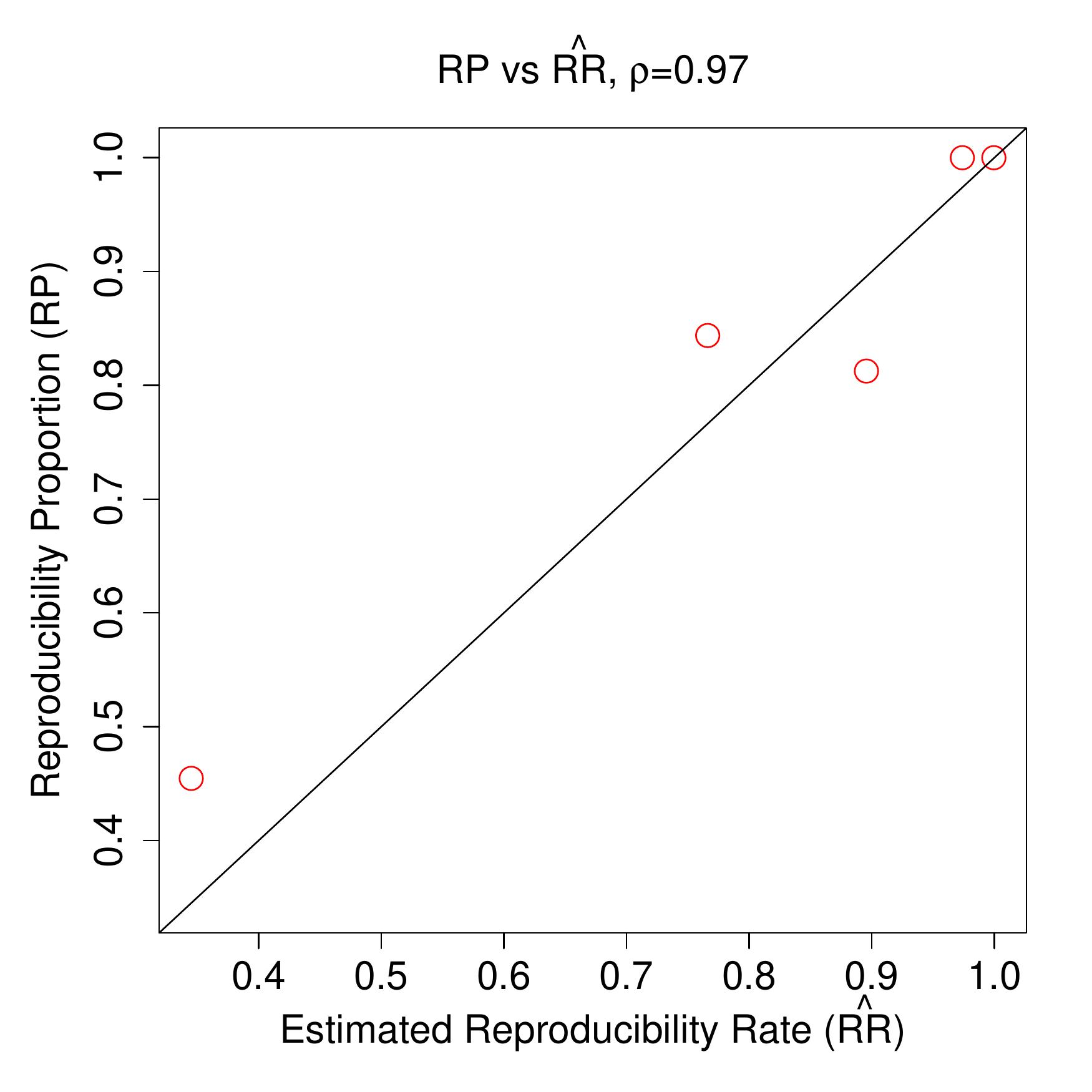}}
    \caption{Reproducibility prediction in LDL Cholesterol data from GLGC. (a) The x-axis is the recall in reproducibility prediction in terms of $\widehat{RR}$, and the y-axis is the corresponding precision. $AUPRC$ is the area under precision-recall curve. Both PR curve based on $\widehat{RR}$ (solid line) and PR curve based on $p$-value (dashed line) are drawn in the figure. According to their $AUPRC$ values, $\widehat{RR}$ predicts reproducibility better than $p$-value. (b) The associations are partitioned into 5 groups according to $\widehat{RR}$. The x-axis is the $\widehat{RR}$ of the group, which is the mid-point of the range of $\widehat{RR}$. The y-axis is the corresponding $RP$ of the group, which is the proportion of the reproduced associations in each group. The solid line is $y=x$.}
\end{figure}

A precision-recall curve is drawn for the prediction of reproduced status based on $\widehat{RR}$ (Figure \ref{PRC_RR_LDL}). The area under the precision-recall curve is 0.968. In comparison, if $p$-value is regarded as an index describing reproducibility, then the area under the curve is 0.919, smaller than the area of $\widehat{RR}$. To see whether $\widehat{RR}$ can describe the reproducibility probability for an association, we calculate $RP$ by partitioning the clumps into 5 groups according to their $\widehat{RR}$. Figure \ref{RP_RR_LDL} shows the good agreement between RP and $\widehat{RR}$ in the partitioned 5 groups of the clumps. The correlation coefficient between them is 0.97.

There are $29$ irreproduced clumps in the result. Their $\widehat{FIR}$ values are all larger than $0.99$, indicating that they have high possibility to be true associations. We carried out meta-analysis again. The corresponding $p$-values are all smaller than $5\times 10^{-8}$.

\subsection{Other potential applications of $RR$}
We can use $RR$ to determine the sample size needed in the replication study to achieve an expected reproducibility probability for the primary associations. Also, we can use $RR$ to check the consistency between the primary study and the replication study when both of them are conducted. We will describe these two potential applications in this subsection.

The following three methods can be used to determine the sample size of a replication study:
\begin{enumerate}
\item Traditional sample size determination method is based on power calculation. A minimum effect size to be detected ($\mu_{min}$) is specified beforehand. Then, the sample size is determined such that the calculated statistical power is larger than a threshold, e.g. $\beta^{(2)}(\mu_{min})>80\%$. This traditional power analysis method treats replication study as another independent primary study. The connection between primary study and replication study is not utilized in the design. The $\mu_{min}$ may be arbitrary. Also bias may occur in the specification of $\mu_{min}$. These make the determined sample size subjective.
\item Taking advantage of the connection between primary study and replication study, the sample size of the replication study can be determined based on calculating Bayesian predictive power $\eta^{(2)}$ \citep{wang2007sample}. But if we consider the major question of replication study, which is how likely the primary positive finding will be replicated, $\eta^{(2)}$ does not directly address this question. For example, $\eta^{(2)}=80\%$ doesn't mean that a primary positive finding has a $80\%$ of chance to be replicated.
\item $RR$ is a comprehensive measure directly addressing the question of replication. For primary study's results, $RR$ is related to $\eta^{(2)}$ in an one-to-one mapping with a better interpretability of ``replication''. If sample size of the replication study is specified with $RR=80\%$, then a primary association has a possibility of $80\%$ to be replicated.
\end{enumerate}
According to above analysis, it is more natural to design a replication study based on $RR$. A very nice property of $RR$ is that we do not really need to carry out the replication study when estimating $RR$. This provides a huge advantage to explore all possibilities when designing the replication study.

Another application of $RR$ is quality check. In normal scenarios, the results of replication study are consistent with $\widehat{RR}$ values. If inconsistency occurs, we should be alarmed. The potential sources of inconsistency should be analyzed. These sources may be attributed to factors influencing either primary study's or replication study's results, such as bias and measurement errors \citep{ioannidis2006non}.

\section{Discussion}\label{discussion}
%[Discussion] Leave some comments about the method\results
%[Introduction-Related work] How to solve the problem in the states of arts, and its shortcomings? (Be generous. "In his inspiring paper [Foo98] Foogle shows...We develop his foundation in the following ways...")
%p value; fdr; power;
Please note that if $fdr^{(1)}>0$, which is usually the case for a primary positive association, $RR$ has an upper limit which is smaller than 1. According to Eq. (\ref{relation}),
\begin{eqnarray}
RR &=& fdr^{(1)} \alpha_2+(1-fdr^{(1)}) \eta^{(2)}\nonumber \\
&\leq & 1-fdr^{(1)}(1-\alpha_2),
\end{eqnarray}
where equality is achieved if $\eta^{(2)}=1$. This indicates that the influence of null distribution (namely $\alpha_2$) never disappears for a primary positive association with $fdr^{(1)}>0$. The Bayesian predictive power $\eta^{(2)}$ can be increased by increasing the sample size of the replication study. In the situation of $fdr^{(1)}>0$, no matter how many individuals are participated in the replication study, the primary association will not have 100\% probability of being reproduced.

Also, since unbiased testing method is used in the replication study, i.e. $\alpha_2\leq \eta^{(2)}$, we have $RR\leq \eta^{(2)}$ according to Eq. (\ref{relation}). This indicates that, for a primary association in a designed replication study, the probability of being reproduced is smaller than its Bayesian predictive power.

At a first glance, people may regard $p$-value as a quantitative index to describe the reproducibility. An association with a lower $p$-value has a higher possibility to be reproduced than an association with a higher $p$-value. The argument is that the $p$-values of associations have the same ordering as the local false discovery rates, which are the probabilities of the corresponding hypotheses being null given their test statistics. But a low probability of being null does not mean a high probability to be reproduced. Hence, unlike $RR$, $p$-value is not an index to describe the reproducibility directly.

%[Discussion] What's the limitation of the method?
% estimation relies on $pi0$ estimation
% LD not be considered
% SE need to be known
% model misspecification
% Other factors in the experiments: confounding(correlation; population stratification (heterogeneity)), measurement errors in genotype/phenotype etc.
%[Discussion] What's the future works may be done?
The accuracy of $RR$ and $FIR$ estimation relies on the accuracy of $\widehat{\pi}_0$. Although we apply the method of \citet{storey2003statistical} to estimate $\pi_0$, there exist other options. For example, when the ``zero assumption'' is violated in data or the true null distribution of test statistics does not agree with the theoretical distribution \citep{efron2004large}, it may be better to use the methods proposed by \citet{langaas2005estimating} or \citet{jin2007estimating} for a reliable estimation of $\pi_0$.

Our model of $RR$ and $FIR$ is limited with independent assumption for each SNP. In reality, the correlation between SNPs, such as linkage disequilibrium, are common. An adjusted model for $RR$ and $FIR$ considering correlation is needed in the future.

\section{Conclusion}\label{conclusion}
% Background theme Recap
%[Conclusion] Briefly recap what you have solved, and its meaning.
% In this paper, we present...
% limitation and future works
In replication based analysis, positive associations identified in the primary study need to be verified in the replication study. In this paper, we presented a Bayesian framework to systematically study the behavior of those primary findings in the replication study, and proposed two new probabilistic measures, reproducibility rate ($RR$) and false irreproducibility rate ($FIR$), to quantify the behavior.

$RR$ is proposed to quantify the reproducibility probability for every finding. An estimation method is provided for $RR$ based on the summary statistics of the primary study. Experiments using simulation and real data show our estimation methods can predict the reproducibility well. Thus, $\widehat{RR}$ can be used to guide the experiments design of the replication study, and can also be used to check whether there are other factors affecting either primary study's or replication study's results, such as bias and measurement errors.

$FIR$ is proposed to quantify the probability that a primary association is still a true positive even when it is not reproduced in the replication study. The irreproduced associations with high $FIR$s also have a high confidence to be truly associated ones. People can use $FIR$ estimation to prevent some irreproduced results from being discarded.

%Limitations:
%\begin{itemize}
%\item Independent assumption is made without considering LD.
%\item SE are assumed to be known without error.
%\end{itemize}

\section*{Acknowledgement}
%[Acknowledge] Warmly acknowledge people who have helped you.
% We thank for... This work was supported with... The program used for...is available from...
This paper was partially supported by a theme-based research project T12-402/13N of the Hong Kong Research Grant Council (RGC). Thanks to the Royal Academy of Engineering for a Research Exchanges with China Award to J.-H. Xue and W. Yu. We thank Prof. D. Donoho at Stanford University for insightful discussions.
\bibliography{R-Rate}

\begin{thebibliography}{}

\bibitem[Balding, 2006]{balding2006tutorial}
Balding, D.~J. (2006).
\newblock A tutorial on statistical methods for population association studies.
\newblock {\em Nature Reviews Genetics}, 7(10):781--791.

\bibitem[Davis and Goadrich, 2006]{davis2006relationship}
Davis, J. and Goadrich, M. (2006).
\newblock The relationship between {Precision-Recall} and {ROC} curves.
\newblock {\em Proceedings of the 23rd international conference on Machine
  learning}, pages 233--240.

\bibitem[Efron, 2004]{efron2004large}
Efron, B. (2004).
\newblock Large-scale simultaneous hypothesis testing: The choice of a null
  hypothesis.
\newblock {\em Journal of the American Statistical Association}, 99:96--104.

\bibitem[Efron, 2005]{efron2005local}
Efron, B. (2005).
\newblock {\em Local false discovery rates}.
\newblock Division of Biostatistics, Stanford University.

\bibitem[Evangelou and Ioannidis, 2013]{evangelou2013meta}
Evangelou, E. and Ioannidis, J.~P. (2013).
\newblock Meta-analysis methods for genome-wide association studies and beyond.
\newblock {\em Nature Reviews Genetics}, 14(6):379--389.

\bibitem[{Global Lipids Genetics Consortium}, 2013]{global2013discovery}
{Global Lipids Genetics Consortium} (2013).
\newblock Discovery and refinement of loci associated with lipid levels.
\newblock {\em Nature Genetics}, 45(11):1274--1283.

\bibitem[Hindorff et~al., rces]{HindorffWebresources}
Hindorff, L.~A., MacArthur, J., Morales, J., Junkins, H.~A., Hall, P.~N.,
  Klemm, A.~K., and Manolio, T.~A. (Web Resources).
\newblock A catalog of published genome-wide association studies.
\newblock {\em Available at: http://www.genome.gov/gwastudies/. Accessed
  [2015.05.28]}.

\bibitem[Hirschhorn and Daly, 2005]{hirschhorn2005genome}
Hirschhorn, J.~N. and Daly, M.~J. (2005).
\newblock Genome-wide association studies for common diseases and complex
  traits.
\newblock {\em Nature Reviews Genetics}, 6(2):95--108.

\bibitem[Ioannidis, 2006]{ioannidis2006non}
Ioannidis, J.~P. (2006).
\newblock Non-replication and inconsistency in the genome-wide association
  setting.
\newblock {\em Human Heredity}, 64(4):203--213.

\bibitem[Jin and Cai, 2007]{jin2007estimating}
Jin, J. and Cai, T.~T. (2007).
\newblock Estimating the null and the proportion of nonnull effects in
  large-scale multiple comparisons.
\newblock {\em Journal of the American Statistical Association},
  102(478):495--506.

\bibitem[Klein et~al., 2005]{klein2005complement}
Klein, R.~J., Zeiss, C., Chew, E.~Y., Tsai, J.-Y., Sackler, R.~S., Haynes, C.,
  Henning, A.~K., SanGiovanni, J.~P., Mane, S.~M., Mayne, S.~T., et~al. (2005).
\newblock Complement factor {H} polymorphism in age-related macular
  degeneration.
\newblock {\em Science}, 308(5720):385--389.

\bibitem[Kraft et~al., 2009]{kraft2009replication}
Kraft, P., Zeggini, E., and Ioannidis, J.~P. (2009).
\newblock Replication in genome-wide association studies.
\newblock {\em Statistical Science}, 24(4):561.

\bibitem[Langaas et~al., 2005]{langaas2005estimating}
Langaas, M., Lindqvist, B.~H., and Ferkingstad, E. (2005).
\newblock Estimating the proportion of true null hypotheses, with application
  to {DNA} microarray data.
\newblock {\em Journal of the Royal Statistical Society: Series B (Statistical
  Methodology)}, 67(4):555--572.

\bibitem[Lecoutre, 2001]{lecoutre2001bayesian}
Lecoutre, B. (2001).
\newblock Bayesian predictive procedure for designing and monitoring
  experiments.
\newblock {\em Bayesian Methods with Applications to Science, Policy and
  Official Statistics, Luxembourg: Office for Official Publications of the
  European Communities}, pages 301--310.

\bibitem[Locke et~al., 2015]{locke2015genetic}
Locke, A.~E., Kahali, B., Berndt, S.~I., Justice, A.~E., Pers, T.~H., Day,
  F.~R., Powell, C., Vedantam, S., Buchkovich, M.~L., Yang, J., et~al. (2015).
\newblock Genetic studies of body mass index yield new insights for obesity
  biology.
\newblock {\em Nature}, 518(7538):197--206.

\bibitem[Morris et~al., 2012]{morris2012large}
Morris, A.~P., Voight, B.~F., Teslovich, T.~M., Ferreira, T., Segr{\'e}, A.~V.,
  Steinthorsdottir, V., Strawbridge, R.~J., Khan, H., Grallert, H., Mahajan,
  A., et~al. (2012).
\newblock Large-scale association analysis provides insights into the genetic
  architecture and pathophysiology of type 2 diabetes.
\newblock {\em Nature Genetics}, 44(9):981.

\bibitem[{NCI-NHGRI Working Group on Replication in Association Studies},
  2007]{chanock2007replicating}
{NCI-NHGRI Working Group on Replication in Association Studies} (2007).
\newblock Replicating genotype--phenotype associations.
\newblock {\em Nature}, 447(7145):655--660.

\bibitem[Park et~al., 2010]{park2010estimation}
Park, J.-H., Wacholder, S., Gail, M.~H., Peters, U., Jacobs, K.~B., Chanock,
  S.~J., and Chatterjee, N. (2010).
\newblock Estimation of effect size distribution from genome-wide association
  studies and implications for future discoveries.
\newblock {\em Nature Genetics}, 42(7):570--575.

\bibitem[Risch and Merikangas, 1996]{risch1996future}
Risch, N. and Merikangas, K. (1996).
\newblock The future of genetic studies of complex human diseases.
\newblock {\em Science}, 273(5281):1516--1517.

\bibitem[Skol et~al., 2006]{skol2006joint}
Skol, A.~D., Scott, L.~J., Abecasis, G.~R., and Boehnke, M. (2006).
\newblock Joint analysis is more efficient than replication-based analysis for
  two-stage genome-wide association studies.
\newblock {\em Nature Genetics}, 38(2):209--213.

\bibitem[Storey and Tibshirani, 2003]{storey2003statistical}
Storey, J.~D. and Tibshirani, R. (2003).
\newblock Statistical significance for genomewide studies.
\newblock {\em Proceedings of the National Academy of Sciences},
  100(16):9440--9445.

\bibitem[Tabor et~al., 2002]{tabor2002candidate}
Tabor, H.~K., Risch, N.~J., and Myers, R.~M. (2002).
\newblock Candidate-gene approaches for studying complex genetic traits:
  practical considerations.
\newblock {\em Nature Reviews Genetics}, 3(5):391--397.

\bibitem[Voight et~al., 2010]{voight2010twelve}
Voight, B.~F., Scott, L.~J., Steinthorsdottir, V., Morris, A.~P., Dina, C.,
  Welch, R.~P., Zeggini, E., Huth, C., Aulchenko, Y.~S., Thorleifsson, G.,
  et~al. (2010).
\newblock Twelve type 2 diabetes susceptibility loci identified through
  large-scale association analysis.
\newblock {\em Nature Genetics}, 42(7):579--589.

\bibitem[Wang, 2007]{wang2007sample}
Wang, M.-D. (2007).
\newblock Sample size reestimation by {Bayesian} prediction.
\newblock {\em Biometrical journal}, 49(3).

\bibitem[Woolf, 1955]{woolf1955estimating}
Woolf, B. (1955).
\newblock On estimating the relation between blood group and disease.
\newblock {\em Ann Hum Genet}, 19(4):251--253.

\bibitem[Yang et~al., 2010]{yang2010common}
Yang, J., Benyamin, B., McEvoy, B.~P., Gordon, S., Henders, A.~K., Nyholt,
  D.~R., Madden, P.~A., Heath, A.~C., Martin, N.~G., Montgomery, G.~W., et~al.
  (2010).
\newblock Common {SNPs} explain a large proportion of the heritability for
  human height.
\newblock {\em Nature Genetics}, 42(7):565--569.

\end{thebibliography}

\clearpage

\appendix
\appendixpage
\begin{appendices}
\section{Detailed deduction of $RR$ and $FIR$}
The relationship between $RR$, $fdr^{(1)}$ and $\beta^{(2)}(\mu)$ can be derived from the law of total probability:
\begin{eqnarray}
RR&=&P(sgn(z^{(1)})Z^{(2)}>z_{\alpha_2} \big| z^{(1)}) \nonumber\\
&=& P(sgn(z^{(1)})Z^{(2)}>z_{\alpha_2},\mathcal{H}_0 \big| z^{(1)})+P(sgn(z^{(1)})Z^{(2)}>z_{\alpha_2},\mathcal{H}_1 \big| z^{(1)}) \nonumber\\
&=& P(\mathcal{H}_0 \big| z^{(1)})P(sgn(z^{(1)})Z^{(2)}>z_{\alpha_2} \big| \mathcal{H}_0,z^{(1)})+P(\mathcal{H}_1 \big| z^{(1)})P(sgn(z^{(1)})Z^{(2)}>z_{\alpha_2} \big| \mathcal{H}_1, z^{(1)}) \nonumber\\
&=& fdr^{(1)}\alpha_2+(1-fdr^{(1)})\eta^{(2)},
\end{eqnarray}
where
\begin{eqnarray}
\eta^{(2)}&=& \int_{-\infty}^{\infty} P(sgn(z^{(1)})Z^{(2)}>z_{\alpha_2}, \mu \big| \mathcal{H}_1, z^{(1)})d\mu \nonumber\\
&=& \int_{-\infty}^{\infty} P(sgn(z^{(1)})Z^{(2)}>z_{\alpha_2} \big| \mathcal{H}_1, \mu, z^{(1)})p(\mu \big| \mathcal{H}_1, z^{(1)})d\mu \nonumber\\
&=& \int_{-\infty}^{\infty} \beta^{(2)}(\mu)p(\mu \big| \mathcal{H}_1, z^{(1)})d\mu \nonumber\\
&=& E(\beta^{(2)}(\mu) \big| z^{(1)}, \mathcal{H}_1).
\end{eqnarray}

The relationship between $FIR$, $fdr^{(1)}$ and $\beta^{(2)}(\mu)$ can be derived using the Bayes formula:
\begin{eqnarray}
FIR&=& P(\mathcal{H}_1 \big| sgn(z^{(1)})Z^{(2)}\leq z_{\alpha_2}, z^{(1)}) \nonumber\\
&=& \frac{P(\mathcal{H}_1 \big| z^{(1)})P(sgn(z^{(1)})Z^{(2)}\leq z_{\alpha_2} \big| \mathcal{H}_1, z^{(1)})}{P(sgn(z^{(1)})Z^{(2)}\leq z_{\alpha_2} \big| z^{(1)})} \nonumber\\
&=& \frac{(1-fdr^{(1)})(1-\eta^{(2)})}{1-RR}.
\end{eqnarray}

\section{Derivation of $fdr^{(1)}$, $\eta^{(2)}$ under a two-component mixture prior}
The following property for multivariate Gaussian distribution can be used to calculate $fdr^{(1)}$ and $\eta^{(2)}$.
\begin{prop}\label{multinormal}
If $\mathbf{Z} \big| \mathbf{\mu}\sim N_p(\mathbf{\mu}, \mathbf{\Sigma})$, and $\mathbf{\mu}\sim N_p(\mathbf{\mu_0}, \mathbf{\Sigma_0})$, then
\begin{eqnarray}
\mathbf{Z}\sim N_p(\mathbf{\mu_0}, \mathbf{\Sigma}+\mathbf{\Sigma_0}) \text{ and } \mathbf{\mu} \big| \mathbf{z}\sim N_p(\mathbf{W\mu_0+(I-W)z}, \mathbf{(I-W)\Sigma})
\end{eqnarray}
with $\mathbf{W}=\Sigma(\Sigma_0+\Sigma)^{-1}$
\end{prop}
The proof of Property \ref{multinormal} can be found in Bishop 2006, Chapter 2.

By using Property \ref{multinormal}, the distribution of the test statistic $Z^{(1)}$ is:
\begin{equation}
Z^{(1)}\sim \pi_0 N(0,1)+(1-\pi_0) N(0, 1+(\frac{\sigma_0}{\sigma^{(1)}})^2).\label{z1dist}
\end{equation}
Hence the local false discovery rate of the primary study can be calculated with following:
\begin{equation}
fdr^{(1)}= \frac{\pi_0 \phi(z^{(1)}) }{\pi_0 \phi(z^{(1)}) +(1-\pi_0) \phi ( \frac{z^{(1)} }{\sqrt{1+(\sigma_0/\sigma^{(1)})^2}} )},
\end{equation}
where $\phi(x)$ is the pdf of the standard normal distribution.

Since $\hat{\mu}^{(j)}\sim N(\mu,(\sigma^{(j)})^2)$ and $(\mu \big| \mathcal{H}_1) \sim N(0, \sigma_0^2)$, we can obtain
\begin{equation}
(\mu \big| z^{(1)},\mathcal{H}_1)\sim N(\lambda \hat{\mu}^{(1)},\lambda (\sigma^{(1)})^2),
\end{equation}
where $\lambda=\frac{1}{1+(\sigma^{(1)}/\sigma_0)^2}$ plays a shrinkage effect. The posterior distribution of $Z^{(2)}$ under $\mathcal{H}_1$ reads
\begin{equation}
(Z^{(2)} \big| z^{(1)},\mathcal{H}_1)\sim N\left(z^*=\lambda \frac{\hat{\mu}^{(1)}}{\sigma^{(2)}}, (\sigma^*)^2=1+\lambda\left(\frac{\sigma^{(1)}}{\sigma^{(2)}}\right)^2 \right).
\end{equation}
The Bayesian predictive power of the replication study can be calculated as follows:
\begin{equation}
\eta^{(2)}= \Phi(\frac{sgn(z^{(1)})z^*-z_{\alpha_2}}{\sigma^*}),
\end{equation}
where $\Phi(x)$ is the cdf of the standard normal distribution.

\subsection*{References}

Bishop, C. M. (2006). Pattern Recognition and Machine Learning. Springer.

\section{Derivation of the $\sigma_0^2$ estimator}
From (\ref{z1dist}), the distribution of $Z^{(1)}$ is a two-component Gaussian mixture model. So we have
\begin{equation}
(Z^{(1)})^2\sim \pi_0 \chi_1^2+(1-\pi_0) \left(1+(\frac{\sigma_0}{\sigma^{(1)}})^2\right)\chi_1^2,
\end{equation}
where $\chi_1^2$ is the $\chi^2$ distribution with degree of freedom (df) 1. The expectation reads
\begin{equation}
E((Z^{(1)})^2)=\pi_0+(1-\pi_0) (1+(\frac{\sigma_0}{\sigma^{(1)}})^2).
\end{equation}
For all SNPs, the following can be obtained
\begin{equation}
E(\sum_{i=1}^m (Z_i^{(1)})^2)=m\pi_0+(1-\pi_0) (m+\sigma_0^2\sum_{i=1}^m (1/\sigma^{(1)}_i)^2).
\end{equation}
By substituting $\sum_{i=1}^m (z_i^{(1)})^2$ for $E(\sum_{i=1}^m (Z_i^{(1)})^2)$, we can get the estimator for $\sigma_0^2$:
\begin{equation}
\hat{\sigma}_0^2=\left(\frac{\sum_{i=1}^m (z^{(1)}_i)^2-m\pi_0}{(1-\pi_0)}-m\right)/\sum_{i=1}^m (1/\sigma^{(1)}_i)^2.
\end{equation}

\clearpage
\section{$\widehat{RR}$ and $\widehat{FIR}$ in T2D data from DIAGRAM}
% latex table generated in R 3.0.1 by xtable 1.7-4 package
% Sun May 31 18:42:45 2015
\begin{table}[ht]
\centering
\caption{$\widehat{RR}$ and $\widehat{FIR}$ results for T2D data in DIAGRAM. Column P1 is the $p$-value in the primary study; Column P2 is the $p$-value in the replication study; Column Pmeta is the $p$-value in the meta-analysis. Column $RR\ 95\%CI$ is the $95\%$ confidence interval for $RR$. Column $FIR\ 95\%CI$ is the $95\%$ confidence interval for $FIR$.}
\begin{tabular}{ccccccccc}
  & SNP & P1 & P2 & Pmeta & RR & RR 95\%CI & FIR & FIR 95\%CI \\
  \hline
1 & rs1801282 & 7.401E-10 & 4.707E-05 & 1.013E-12 & 0.779 & (0.769 , 0.788) & 1.000 & (1.000 , 1.000) \\
  2 & rs11711477 & 1.339E-11 & 6.166E-08 & 2.151E-17 & 0.951 & (0.949 , 0.953) & 1.000 & (1.000 , 1.000) \\
  3 & rs1801214 & 5.073E-10 & 2.887E-07 & 3.149E-15 & 0.913 & (0.910 , 0.916) & 1.000 & (1.000 , 1.000) \\
  4 & rs9348440 & 2.859E-12 & 1.369E-06 & 1.638E-16 & 0.923 & (0.918 , 0.928) & 1.000 & (1.000 , 1.000) \\
  5 & rs4710940 & 3.553E-15 & 9.980E-09 & 2.212E-21 & 0.991 & (0.991 , 0.992) & 1.000 & (1.000 , 1.000) \\
  6 & rs6931514 & 0.000E+00 & 1.342E-13 & 5.207E-32 & 1.000 & (1.000 , 1.000) & 1.000 & (1.000 , 1.000) \\
  7 & rs9465871 & 0.000E+00 & 4.163E-11 & 6.427E-26 & 1.000 & (1.000 , 1.000) & 1.000 & (1.000 , 1.000) \\
  8 & rs7741604 & 4.868E-09 & 1.750E-03 & 3.803E-10 & 0.680 & (0.670 , 0.689) & 1.000 & (1.000 , 1.000) \\
  9 & rs864745 & 1.506E-12 & 3.423E-06 & 2.165E-16 & 0.938 & (0.936 , 0.940) & 1.000 & (1.000 , 1.000) \\
  10 & rs498475 & 1.398E-08 & 6.537E-07 & 1.568E-13 & 0.866 & (0.862 , 0.870) & 1.000 & (1.000 , 1.000) \\
  11 & rs11774700 & 5.053E-11 & 1.251E-07 & 1.414E-15 & 0.993 & (0.992 , 0.993) & 1.000 & (1.000 , 1.000) \\
  12 & rs10811661 & 1.976E-14 & 1.210E-14 & 3.450E-27 & 0.953 & (0.950 , 0.956) & 1.000 & (1.000 , 1.000) \\
  13 & rs2798253 & 3.249E-12 & 2.282E-04 & 5.610E-14 & 0.863 & (0.858 , 0.869) & 1.000 & (1.000 , 1.000) \\
  14 & rs2421943 & 5.073E-10 & 8.855E-01 & 6.072E-09 & 0.001 & (0.001 , 0.001) & 1.000 & (1.000 , 1.000) \\
  15 & rs7911264 & 1.339E-11 & 4.413E-07 & 5.623E-16 & 0.984 & (0.983 , 0.985) & 1.000 & (1.000 , 1.000) \\
  16 & rs7923866 & 2.551E-13 & 5.355E-08 & 1.126E-18 & 0.990 & (0.989 , 0.990) & 1.000 & (1.000 , 1.000) \\
  17 & rs7917983 & 0.000E+00 & 1.442E-08 & 2.478E-27 & 0.998 & (0.998 , 0.999) & 1.000 & (1.000 , 1.000) \\
  18 & rs10128255 & 2.854E-08 & 5.551E-16 & 2.103E-22 & 0.923 & (0.919 , 0.926) & 1.000 & (1.000 , 1.000) \\
  19 & rs12266632 & 9.452E-11 & 1.256E-10 & 2.241E-19 & 0.876 & (0.861 , 0.891) & 1.000 & (1.000 , 1.000) \\
  20 & rs10787472 & 0.000E+00 & 0.000E+00 & 1.336E-58 & 1.000 & (1.000 , 1.000) & 1.000 & (1.000 , 1.000) \\
  21 & rs12255372 & 0.000E+00 & 0.000E+00 & 2.537E-112 & 1.000 & (1.000 , 1.000) & 1.000 & (1.000 , 1.000) \\
  22 & rs11196212 & 7.313E-11 & 1.306E-12 & 1.326E-21 & 0.880 & (0.877 , 0.883) & 1.000 & (1.000 , 1.000) \\
  23 & rs10765573 & 5.073E-10 & 6.081E-04 & 3.363E-11 & 0.888 & (0.884 , 0.892) & 1.000 & (1.000 , 1.000) \\
  24 & rs12149832 & 1.339E-11 & 3.622E-12 & 6.775E-22 & 0.967 & (0.966 , 0.969) & 1.000 & (1.000 , 1.000) \\
  \end{tabular}
\end{table}

\section{$\widehat{RR}$ and $\widehat{FIR}$ in LDL Cholesterol data from GLGC}
% latex table generated in R 3.0.1 by xtable 1.7-4 package
% Sun May 31 18:41:14 2015
\begin{center}
\begin{longtable}{ccccccccc}
 \caption{$\widehat{RR}$ and $\widehat{FIR}$ results for LDL Cholesterol data in GLGC. Column P1 is the $p$-value in the primary study; Column P2 is the $p$-value in the replication study; Column Pmeta is the $p$-value in the meta-analysis. Column $RR\ 95\%CI$ is the $95\%$ confidence interval for $RR$. Column $FIR\ 95\%CI$ is the $95\%$ confidence interval for $FIR$. When $\widehat{RR}=1$, $\widehat{FIR}$ cannot be obtained.}\\
  & SNP & P1 & P2 & Pmeta & RR & RR 95\%CI & FIR & FIR 95\%CI \\
  \hline
1 & rs2304130 & 0.000E+00 & 3.236E-12 & 1.004E-34 & 0.999 & (0.999 , 1.000) & 1.000 & (1.000 , 1.000) \\
  2 & rs7832643 & 7.963E-09 & 2.739E-12 & 4.619E-19 & 0.709 & (0.705 , 0.714) & 1.000 & (1.000 , 1.000) \\
  3 & rs10808546 & 0.000E+00 & 0.000E+00 & 1.475E-47 & 1.000 & (1.000 , 1.000) & 1.000 & (1.000 , 1.000) \\
  4 & rs7515901 & 4.420E-09 & 1.166E-14 & 5.260E-21 & 0.557 & (0.549 , 0.566) & 1.000 & (1.000 , 1.000) \\
  5 & rs10069744 & 3.650E-10 & 3.165E-05 & 3.143E-13 & 0.779 & (0.769 , 0.789) & 1.000 & (1.000 , 1.000) \\
  6 & rs13344893 & 1.510E-14 & 0.000E+00 & 5.083E-30 & 0.953 & (0.951 , 0.956) & 1.000 & (1.000 , 1.000) \\
  7 & rs6725189 & 1.332E-15 & 0.000E+00 & 4.481E-41 & 0.984 & (0.983 , 0.985) & 1.000 & (1.000 , 1.000) \\
  8 & rs16996148 & 0.000E+00 & 0.000E+00 & 5.057E-49 & 1.000 & (1.000 , 1.000) & 1.000 & (1.000 , 1.000) \\
  9 & rs7251031 & 7.840E-13 & 3.252E-12 & 3.653E-23 & 0.869 & (0.865 , 0.874) & 1.000 & (1.000 , 1.000) \\
  10 & rs12286037 & 1.658E-11 & 1.130E-10 & 2.245E-20 & 0.781 & (0.766 , 0.797) & 1.000 & (1.000 , 1.000) \\
  11 & rs631106 & 0.000E+00 & 0.000E+00 & 3.108E-34 & 0.993 & (0.993 , 0.993) & 1.000 & (1.000 , 1.000) \\
  12 & rs3791981 & 0.000E+00 & 0.000E+00 & 1.263E-44 & 0.999 & (0.999 , 1.000) & 1.000 & (1.000 , 1.000) \\
  13 & rs2385114 & 5.646E-10 & 4.558E-12 & 3.413E-20 & 0.853 & (0.849 , 0.857) & 1.000 & (1.000 , 1.000) \\
  14 & rs3810444 & 1.024E-12 & 1.049E-03 & 1.098E-13 & 0.690 & (0.671 , 0.710) & 1.000 & (1.000 , 1.000) \\
  15 & rs10422616 & 5.646E-09 & 4.712E-07 & 2.843E-14 & 0.722 & (0.718 , 0.727) & 1.000 & (1.000 , 1.000) \\
  16 & rs4518686 & 9.917E-09 & 0.000E+00 & 1.700E-25 & 0.802 & (0.798 , 0.807) & 1.000 & (1.000 , 1.000) \\
  17 & rs10062361 & 0.000E+00 & 0.000E+00 & 9.229E-59 & 1.000 & (1.000 , 1.000) & 1.000 & (1.000 , 1.000) \\
  18 & rs12708967 & 4.138E-08 & 1.777E-05 & 6.755E-12 & 0.435 & (0.428 , 0.443) & 1.000 & (1.000 , 1.000) \\
  19 & rs2479409 & 0.000E+00 & 0.000E+00 & 2.905E-55 & 1.000 & (1.000 , 1.000) & 1.000 & (1.000 , 1.000) \\
  20 & rs413582 & 4.467E-12 & 0.000E+00 & 2.488E-32 & 0.916 & (0.914 , 0.919) & 1.000 & (1.000 , 1.000) \\
  21 & rs2073547 & 1.843E-14 & 1.777E-10 & 4.249E-23 & 0.826 & (0.821 , 0.832) & 1.000 & (1.000 , 1.000) \\
  22 & rs655246 & 0.000E+00 & 0.000E+00 & 2.545E-46 & 0.999 & (0.999 , 0.999) & 1.000 & (1.000 , 1.000) \\
  23 & rs10198175 & 3.553E-15 & 0.000E+00 & 3.494E-34 & 0.944 & (0.940 , 0.949) & 1.000 & (1.000 , 1.000) \\
  24 & rs2954038 & 8.882E-16 & 3.191E-12 & 4.043E-26 & 0.961 & (0.960 , 0.963) & 1.000 & (1.000 , 1.000) \\
  25 & rs2737252 & 1.070E-08 & 1.677E-07 & 1.881E-14 & 0.728 & (0.722 , 0.733) & 1.000 & (1.000 , 1.000) \\
  26 & rs4703646 & 0.000E+00 & 0.000E+00 & 5.373E-35 & 0.997 & (0.997 , 0.997) & 1.000 & (1.000 , 1.000) \\
  27 & rs2642438 & 6.851E-10 & 6.737E-09 & 5.250E-17 & 0.834 & (0.830 , 0.839) & 1.000 & (1.000 , 1.000) \\
  28 & rs9302635 & 2.070E-08 & 1.694E-09 & 4.569E-16 & 0.592 & (0.585 , 0.600) & 1.000 & (1.000 , 1.000) \\
  29 & rs17584208 & 0.000E+00 & 0.000E+00 & 1.513E-51 & 0.996 & (0.995 , 0.997) & 1.000 & (1.000 , 1.000) \\
  30 & rs4803750 & 0.000E+00 & 0.000E+00 & 2.688E-173 & 1.000 & (1.000 , 1.000) &   &   \\
  31 & rs688 & 0.000E+00 & 0.000E+00 & 3.040E-48 & 0.998 & (0.997 , 0.998) & 1.000 & (1.000 , 1.000) \\
  32 & rs12410656 & 2.197E-08 & 3.282E-03 & 5.072E-10 & 0.052 & (0.050 , 0.055) & 1.000 & (1.000 , 1.000) \\
  33 & rs8044335 & 4.509E-10 & 2.607E-08 & 1.337E-16 & 0.822 & (0.818 , 0.826) & 1.000 & (1.000 , 1.000) \\
  34 & rs5744680 & 0.000E+00 & 0.000E+00 & 7.507E-68 & 1.000 & (1.000 , 1.000) & 1.000 & (1.000 , 1.000) \\
  35 & rs7523242 & 0.000E+00 & 0.000E+00 & 2.067E-45 & 0.999 & (0.998 , 0.999) & 1.000 & (1.000 , 1.000) \\
  36 & rs6129778 & 6.113E-10 & 1.065E-08 & 7.860E-17 & 0.850 & (0.845 , 0.855) & 1.000 & (1.000 , 1.000) \\
  37 & rs10832962 & 5.156E-10 & 1.804E-07 & 1.244E-15 & 0.841 & (0.837 , 0.845) & 1.000 & (1.000 , 1.000) \\
  38 & rs10947332 & 3.798E-08 & 1.985E-13 & 2.257E-19 & 0.626 & (0.616 , 0.637) & 1.000 & (1.000 , 1.000) \\
  39 & rs622342 & 7.061E-09 & 1.530E-08 & 1.244E-15 & 0.842 & (0.838 , 0.846) & 1.000 & (1.000 , 1.000) \\
  40 & rs2288912 & 3.600E-08 & 7.003E-03 & 1.078E-08 & 0.626 & (0.621 , 0.631) & 1.000 & (1.000 , 1.000) \\
  41 & rs3786721 & 1.754E-14 & 0.000E+00 & 7.448E-35 & 0.949 & (0.947 , 0.950) & 1.000 & (1.000 , 1.000) \\
  42 & rs693 & 0.000E+00 & 0.000E+00 & 4.357E-139 & 1.000 & (1.000 , 1.000) & 1.000 & (1.000 , 1.000) \\
  43 & rs505151 & 3.798E-08 & 7.804E-12 & 4.718E-18 & 0.431 & (0.405 , 0.459) & 1.000 & (1.000 , 1.000) \\
  44 & rs11881156 & 0.000E+00 & 0.000E+00 & 1.574E-61 & 1.000 & (1.000 , 1.000) & 1.000 & (1.000 , 1.000) \\
  45 & rs174570 & 9.241E-11 & 2.842E-14 & 4.965E-23 & 0.845 & (0.839 , 0.851) & 1.000 & (1.000 , 1.000) \\
  46 & rs6756629 & 0.000E+00 & 3.142E-14 & 3.410E-50 & 0.920 & (0.910 , 0.929) & 1.000 & (1.000 , 1.000) \\
  47 & rs4962153 & 3.843E-10 & 1.033E-05 & 3.893E-14 & 0.237 & (0.230 , 0.244) & 1.000 & (1.000 , 1.000) \\
  48 & rs10402271 & 0.000E+00 & 0.000E+00 & 1.945E-131 & 1.000 & (1.000 , 1.000) &  &   \\
  49 & rs10903129 & 2.399E-10 & 2.745E-10 & 7.660E-19 & 0.838 & (0.835 , 0.842) & 1.000 & (1.000 , 1.000) \\
  50 & rs7205804 & 0.000E+00 & 1.640E-09 & 2.174E-27 & 0.996 & (0.996 , 0.996) & 1.000 & (1.000 , 1.000) \\
  51 & rs2738464 & 2.603E-08 & 1.529E-05 & 4.580E-12 & 0.561 & (0.550 , 0.574) & 1.000 & (1.000 , 1.000) \\
  52 & rs1000237 & 4.936E-12 & 2.317E-04 & 1.261E-13 & 0.917 & (0.915 , 0.920) & 1.000 & (1.000 , 1.000) \\
  53 & rs6662286 & 0.000E+00 & 9.770E-15 & 8.148E-42 & 0.982 & (0.979 , 0.984) & 1.000 & (1.000 , 1.000) \\
  54 & rs4360309 & 1.549E-09 & 8.146E-13 & 3.245E-20 & 0.719 & (0.715 , 0.724) & 1.000 & (1.000 , 1.000) \\
  55 & rs6065311 & 2.442E-15 & 0.000E+00 & 5.006E-31 & 0.989 & (0.989 , 0.990) & 1.000 & (1.000 , 1.000) \\
  56 & rs3786722 & 0.000E+00 & 0.000E+00 & 7.761E-69 & 1.000 & (1.000 , 1.000) & 1.000 & (1.000 , 1.000) \\
  57 & rs11668536 & 3.553E-15 & 0.000E+00 & 4.181E-30 & 0.948 & (0.946 , 0.950) & 1.000 & (1.000 , 1.000) \\
  58 & rs17135399 & 4.997E-12 & 7.671E-12 & 7.810E-22 & 0.573 & (0.554 , 0.593) & 1.000 & (1.000 , 1.000) \\
  59 & rs7552841 & 3.105E-11 & 1.593E-07 & 6.081E-17 & 0.450 & (0.445 , 0.456) & 1.000 & (1.000 , 1.000) \\
  60 & rs7030248 & 8.480E-12 & 5.626E-05 & 2.155E-14 & 0.882 & (0.879 , 0.885) & 1.000 & (1.000 , 1.000) \\
  61 & rs6589566 & 6.037E-12 & 2.220E-16 & 2.413E-26 & 0.843 & (0.831 , 0.855) & 1.000 & (1.000 , 1.000) \\
  62 & rs752434 & 5.536E-09 & 2.403E-11 & 2.258E-18 & 0.703 & (0.698 , 0.708) & 1.000 & (1.000 , 1.000) \\
  63 & rs10401969 & 0.000E+00 & 0.000E+00 & 9.192E-61 & 1.000 & (1.000 , 1.000) & 1.000 & (1.000 , 1.000) \\
  64 & rs4341893 & 0.000E+00 & 0.000E+00 & 1.594E-56 & 1.000 & (1.000 , 1.000) & 1.000 & (1.000 , 1.000) \\
  65 & rs540796 & 0.000E+00 & 1.793E-13 & 7.633E-37 & 1.000 & (1.000 , 1.000) & 1.000 & (1.000 , 1.000) \\
  66 & rs4426495 & 0.000E+00 & 2.109E-15 & 6.043E-32 & 0.997 & (0.996 , 0.997) & 1.000 & (1.000 , 1.000) \\
  67 & rs568938 & 0.000E+00 & 0.000E+00 & 7.217E-155 & 1.000 & (1.000 , 1.000) &  &   \\
  68 & rs6739502 & 1.973E-09 & 0.000E+00 & 8.875E-25 & 0.834 & (0.831 , 0.838) & 1.000 & (1.000 , 1.000) \\
  69 & rs2244608 & 2.477E-09 & 4.140E-13 & 2.114E-20 & 0.776 & (0.771 , 0.780) & 1.000 & (1.000 , 1.000) \\
  70 & rs17424122 & 5.068E-10 & 3.203E-04 & 1.362E-12 & 0.132 & (0.125 , 0.141) & 1.000 & (1.000 , 1.000) \\
  71 & rs769450 & 3.135E-13 & 3.649E-01 & 5.366E-13 & 0.001 & (0.001 , 0.002) & 1.000 & (1.000 , 1.000) \\
  72 & rs4299376 & 0.000E+00 & 0.000E+00 & 8.724E-73 & 1.000 & (1.000 , 1.000) & 1.000 & (1.000 , 1.000) \\
  73 & rs10403668 & 2.380E-11 & 8.471E-08 & 3.111E-17 & 0.863 & (0.857 , 0.869) & 1.000 & (1.000 , 1.000) \\
  74 & rs174448 & 3.528E-11 & 5.276E-05 & 1.668E-14 & 0.245 & (0.241 , 0.249) & 1.000 & (1.000 , 1.000) \\
  75 & rs7701925 & 0.000E+00 & 6.267E-04 & 6.506E-27 & 0.113 & (0.111 , 0.116) & 1.000 & (1.000 , 1.000) \\
  76 & rs6722374 & 6.555E-10 & 2.585E-03 & 1.500E-10 & 0.812 & (0.808 , 0.815) & 1.000 & (1.000 , 1.000) \\
  77 & rs17398765 & 6.373E-14 & 0.000E+00 & 1.879E-33 & 0.848 & (0.835 , 0.860) & 1.000 & (1.000 , 1.000) \\
  78 & rs16973520 & 2.931E-11 & 1.825E-11 & 6.538E-21 & 0.922 & (0.919 , 0.925) & 1.000 & (1.000 , 1.000) \\
  79 & rs174601 & 0.000E+00 & 0.000E+00 & 3.385E-37 & 0.992 & (0.992 , 0.993) & 1.000 & (1.000 , 1.000) \\
  80 & rs3106167 & 4.516E-12 & 1.013E-05 & 2.665E-15 & 0.919 & (0.914 , 0.924) & 1.000 & (1.000 , 1.000) \\
  81 & rs12710745 & 1.530E-13 & 0.000E+00 & 1.479E-29 & 0.947 & (0.945 , 0.949) & 1.000 & (1.000 , 1.000) \\
  82 & rs4722551 & 1.797E-09 & 7.813E-08 & 1.468E-15 & 0.689 & (0.682 , 0.697) & 1.000 & (1.000 , 1.000) \\
  83 & rs10515198 & 6.943E-12 & 1.015E-12 & 9.264E-23 & 0.872 & (0.865 , 0.880) & 1.000 & (1.000 , 1.000) \\
  84 & rs4808199 & 0.000E+00 & 1.193E-06 & 8.426E-21 & 0.991 & (0.991 , 0.992) & 1.000 & (1.000 , 1.000) \\
  85 & rs4507059 & 3.987E-10 & 1.301E-11 & 7.174E-20 & 0.813 & (0.809 , 0.818) & 1.000 & (1.000 , 1.000) \\
  86 & rs3903032 & 5.730E-10 & 1.665E-15 & 2.501E-23 & 0.786 & (0.776 , 0.796) & 1.000 & (1.000 , 1.000) \\
  87 & rs13465 & 0.000E+00 & 1.110E-16 & 8.517E-34 & 0.905 & (0.894 , 0.916) & 1.000 & (1.000 , 1.000) \\
  88 & rs12608822 & 1.220E-11 & 1.328E-05 & 1.914E-15 & 0.401 & (0.381 , 0.423) & 1.000 & (1.000 , 1.000) \\
  89 & rs912540 & 1.191E-11 & 3.157E-12 & 4.812E-22 & 0.897 & (0.893 , 0.902) & 1.000 & (1.000 , 1.000) \\
  90 & rs1168114 & 0.000E+00 & 0.000E+00 & 3.326E-36 & 0.998 & (0.998 , 0.998) & 1.000 & (1.000 , 1.000) \\
  91 & rs17150482 & 3.100E-08 & 4.200E-05 & 1.677E-11 & 0.640 & (0.635 , 0.646) & 1.000 & (1.000 , 1.000) \\
  92 & rs6511720 & 0.000E+00 & 0.000E+00 & 3.785E-287 & 1.000 & (1.000 , 1.000) &  &   \\
  93 & rs934197 & 0.000E+00 & 1.667E-03 & 4.659E-84 & 0.195 & (0.191 , 0.199) & 1.000 & (1.000 , 1.000) \\
  94 & rs10198972 & 2.705E-12 & 5.551E-16 & 4.175E-26 & 0.692 & (0.668 , 0.718) & 1.000 & (1.000 , 1.000) \\
  95 & rs9989419 & 6.937E-09 & 1.368E-05 & 1.225E-12 & 0.694 & (0.689 , 0.699) & 1.000 & (1.000 , 1.000) \\
  96 & rs10893500 & 1.815E-09 & 2.209E-14 & 8.298E-22 & 0.801 & (0.794 , 0.809) & 1.000 & (1.000 , 1.000) \\
  97 & rs635634 & 0.000E+00 & 1.488E-14 & 9.338E-45 & 0.993 & (0.992 , 0.993) & 1.000 & (1.000 , 1.000) \\
  98 & rs6711016 & 0.000E+00 & 0.000E+00 & 1.370E-37 & 0.991 & (0.990 , 0.991) & 1.000 & (1.000 , 1.000) \\
  99 & rs12448528 & 8.317E-09 & 1.461E-05 & 1.116E-12 & 0.276 & (0.271 , 0.282) & 1.000 & (1.000 , 1.000) \\
  100 & rs4077440 & 7.394E-14 & 1.079E-11 & 1.160E-23 & 0.972 & (0.971 , 0.973) & 1.000 & (1.000 , 1.000) \\
  101 & rs2972564 & 1.599E-11 & 6.825E-13 & 2.846E-22 & 0.655 & (0.642 , 0.669) & 1.000 & (1.000 , 1.000) \\
  102 & rs10888897 & 0.000E+00 & 1.407E-10 & 1.495E-33 & 0.995 & (0.995 , 0.995) & 1.000 & (1.000 , 1.000) \\
  103 & rs8108762 & 2.321E-08 & 5.551E-16 & 7.272E-22 & 0.737 & (0.732 , 0.742) & 1.000 & (1.000 , 1.000) \\
  104 & rs2278444 & 0.000E+00 & 0.000E+00 & 1.640E-37 & 0.997 & (0.997 , 0.997) & 1.000 & (1.000 , 1.000) \\
  105 & rs12294259 & 2.917E-12 & 7.115E-12 & 2.719E-22 & 0.879 & (0.867 , 0.891) & 1.000 & (1.000 , 1.000) \\
  106 & rs4148218 & 5.875E-11 & 7.889E-12 & 6.413E-21 & 0.826 & (0.820 , 0.832) & 1.000 & (1.000 , 1.000) \\
  107 & rs7571647 & 0.000E+00 & 3.413E-11 & 1.880E-43 & 0.945 & (0.940 , 0.950) & 1.000 & (1.000 , 1.000) \\
  108 & rs17231506 & 0.000E+00 & 1.200E-04 & 9.691E-33 & 0.940 & (0.937 , 0.942) & 1.000 & (1.000 , 1.000) \\
  109 & rs4738684 & 9.671E-12 & 3.481E-09 & 4.619E-19 & 0.918 & (0.915 , 0.920) & 1.000 & (1.000 , 1.000) \\
  110 & rs9305020 & 0.000E+00 & 0.000E+00 & 3.579E-167 & 1.000 & (1.000 , 1.000) &  &   \\
  111 & rs17034539 & 9.859E-14 & 6.588E-13 & 8.673E-25 & 0.956 & (0.953 , 0.958) & 1.000 & (1.000 , 1.000) \\
  112 & rs6016381 & 2.220E-16 & 7.925E-08 & 1.264E-21 & 0.990 & (0.990 , 0.991) & 1.000 & (1.000 , 1.000) \\
  113 & rs4530754 & 1.549E-09 & 9.395E-07 & 2.191E-14 & 0.841 & (0.838 , 0.844) & 1.000 & (1.000 , 1.000) \\
  114 & rs16979372 & 0.000E+00 & 0.000E+00 & 6.191E-52 & 0.993 & (0.991 , 0.995) & 1.000 & (1.000 , 1.000) \\
  115 & rs8176720 & 1.760E-09 & 1.180E-10 & 2.397E-18 & 0.838 & (0.835 , 0.842) & 1.000 & (1.000 , 1.000) \\
  116 & rs4927207 & 0.000E+00 & 0.000E+00 & 2.762E-45 & 1.000 & (1.000 , 1.000) & 1.000 & (1.000 , 1.000) \\
  117 & rs6511727 & 7.586E-12 & 1.719E-03 & 2.560E-12 & 0.897 & (0.894 , 0.899) & 1.000 & (1.000 , 1.000) \\
  118 & rs2479408 & 0.000E+00 & 2.261E-04 & 8.723E-27 & 0.871 & (0.866 , 0.877) & 1.000 & (1.000 , 1.000) \\
  119 & rs531819 & 0.000E+00 & 0.000E+00 & 4.421E-147 & 1.000 & (1.000 , 1.000) & 1.000 & (1.000 , 1.000) \\
  120 & rs13420469 & 0.000E+00 & 0.000E+00 & 2.393E-60 & 1.000 & (1.000 , 1.000) & 1.000 & (1.000 , 1.000) \\
  121 & rs588245 & 1.157E-08 & 3.121E-04 & 3.044E-11 & 0.092 & (0.091 , 0.094) & 1.000 & (1.000 , 1.000) \\
  122 & rs688386 & 0.000E+00 & 0.000E+00 & 2.279E-44 & 0.999 & (0.999 , 0.999) & 1.000 & (1.000 , 1.000) \\
  123 & rs6878680 & 1.855E-11 & 0.000E+00 & 2.364E-32 & 0.908 & (0.905 , 0.910) & 1.000 & (1.000 , 1.000) \\
  124 & rs11753995 & 3.357E-08 & 2.220E-16 & 7.994E-22 & 0.633 & (0.625 , 0.641) & 1.000 & (1.000 , 1.000) \\
  125 & rs12748152 & 2.111E-09 & 1.487E-06 & 4.012E-14 & 0.672 & (0.659 , 0.686) & 1.000 & (1.000 , 1.000) \\
  126 & rs11244041 & 1.683E-08 & 5.281E-01 & 2.932E-08 & 0.000 & (0.000 , 0.000) & 1.000 & (1.000 , 1.000) \\
  127 & rs1864163 & 0.000E+00 & 9.250E-08 & 2.704E-22 & 0.973 & (0.972 , 0.975) & 1.000 & (1.000 , 1.000) \\
  128 & rs10495907 & 2.192E-10 & 1.576E-06 & 5.824E-15 & 0.810 & (0.803 , 0.817) & 1.000 & (1.000 , 1.000) \\
  129 & rs4926670 & 0.000E+00 & 0.000E+00 & 1.162E-45 & 1.000 & (1.000 , 1.000) & 1.000 & (1.000 , 1.000) \\
  130 & rs6729410 & 2.488E-09 & 7.772E-16 & 2.359E-22 & 0.660 & (0.655 , 0.665) & 1.000 & (1.000 , 1.000) \\
  131 & rs2194562 & 4.697E-10 & 8.657E-06 & 6.793E-14 & 0.715 & (0.704 , 0.727) & 1.000 & (1.000 , 1.000) \\
  132 & rs461473 & 2.648E-08 & 7.008E-05 & 2.526E-11 & 0.545 & (0.532 , 0.558) & 1.000 & (1.000 , 1.000) \\
  133 & rs157580 & 0.000E+00 & 0.000E+00 & 8.087E-127 & 1.000 & (1.000 , 1.000) & 1.000 & (1.000 , 1.000) \\
  134 & rs9929488 & 3.502E-08 & 7.434E-08 & 2.694E-14 & 0.715 & (0.708 , 0.723) & 1.000 & (1.000 , 1.000) \\
  135 & rs2524299 & 6.356E-12 & 6.704E-08 & 5.892E-18 & 0.841 & (0.835 , 0.848) & 1.000 & (1.000 , 1.000) \\
  136 & rs12677676 & 0.000E+00 & 3.671E-02 & 7.490E-18 & 0.109 & (0.107 , 0.112) & 1.000 & (1.000 , 1.000) \\
  137 & rs9293656 & 0.000E+00 & 0.000E+00 & 5.423E-44 & 0.999 & (0.999 , 0.999) & 1.000 & (1.000 , 1.000) \\
  138 & rs7551981 & 0.000E+00 & 0.000E+00 & 2.009E-35 & 0.996 & (0.996 , 0.997) & 1.000 & (1.000 , 1.000) \\
  139 & rs11206551 & 6.373E-14 & 1.963E-11 & 1.917E-23 & 0.969 & (0.968 , 0.971) & 1.000 & (1.000 , 1.000) \\
  140 & rs17800760 & 9.193E-14 & 2.688E-10 & 3.694E-22 & 0.938 & (0.935 , 0.942) & 1.000 & (1.000 , 1.000) \\
  141 & rs253412 & 0.000E+00 & 0.000E+00 & 5.339E-42 & 0.996 & (0.996 , 0.997) & 1.000 & (1.000 , 1.000) \\
  142 & rs2075650 & 0.000E+00 & 0.000E+00 & 1.835E-226 & 1.000 & (1.000 , 1.000) &  &   \\
  143 & rs17646665 & 1.437E-11 & 1.099E-14 & 2.085E-24 & 0.870 & (0.856 , 0.884) & 1.000 & (1.000 , 1.000) \\
  144 & rs8103315 & 0.000E+00 & 1.143E-07 & 1.249E-22 & 0.920 & (0.915 , 0.925) & 1.000 & (1.000 , 1.000) \\
  145 & rs217386 & 1.926E-11 & 4.911E-12 & 1.264E-21 & 0.880 & (0.877 , 0.883) & 1.000 & (1.000 , 1.000) \\
  146 & rs1175544 & 1.399E-09 & 2.120E-05 & 5.799E-13 & 0.794 & (0.789 , 0.798) & 1.000 & (1.000 , 1.000) \\
  147 & rs4240624 & 1.084E-13 & 8.327E-15 & 1.458E-26 & 0.906 & (0.899 , 0.913) & 1.000 & (1.000 , 1.000) \\
  148 & rs6547409 & 0.000E+00 & 0.000E+00 & 4.256E-45 & 0.923 & (0.913 , 0.932) & 1.000 & (1.000 , 1.000) \\
  149 & rs6859 & 0.000E+00 & 0.000E+00 & 1.072E-101 & 1.000 & (1.000 , 1.000) & 1.000 & (1.000 , 1.000) \\
  150 & rs4704200 & 0.000E+00 & 0.000E+00 & 7.507E-68 & 1.000 & (1.000 , 1.000) & 1.000 & (1.000 , 1.000) \\
  151 & rs2297374 & 6.541E-13 & 5.434E-07 & 1.203E-17 & 0.950 & (0.948 , 0.952) & 1.000 & (1.000 , 1.000) \\
  152 & rs2000999 & 0.000E+00 & 0.000E+00 & 2.466E-45 & 1.000 & (1.000 , 1.000) & 1.000 & (1.000 , 1.000) \\
  153 & rs754524 & 0.000E+00 & 0.000E+00 & 2.321E-116 & 1.000 & (1.000 , 1.000) & 1.000 & (1.000 , 1.000) \\
  154 & rs8104483 & 2.220E-16 & 0.000E+00 & 8.050E-35 & 0.992 & (0.991 , 0.992) & 1.000 & (1.000 , 1.000) \\
  155 & rs8106664 & 0.000E+00 & 2.043E-14 & 1.025E-30 & 0.995 & (0.994 , 0.995) & 1.000 & (1.000 , 1.000) \\
  156 & rs413380 & 1.637E-09 & 6.317E-11 & 1.554E-18 & 0.448 & (0.423 , 0.474) & 1.000 & (1.000 , 1.000) \\
  157 & rs17397667 & 0.000E+00 & 0.000E+00 & 4.128E-57 & 1.000 & (1.000 , 1.000) & 1.000 & (1.000 , 1.000) \\
  158 & rs12720804 & 2.328E-10 & 4.194E-06 & 9.497E-15 & 0.002 & (0.002 , 0.003) & 0.982 & (0.976 , 0.988) \\
  159 & rs11124924 & 2.229E-09 & 2.556E-08 & 6.302E-16 & 0.681 & (0.665 , 0.697) & 1.000 & (1.000 , 1.000) \\
  160 & rs714948 & 6.974E-10 & 4.339E-06 & 2.890E-14 & 0.397 & (0.387 , 0.408) & 1.000 & (1.000 , 1.000) \\
  161 & rs11206510 & 0.000E+00 & 0.000E+00 & 4.993E-62 & 1.000 & (1.000 , 1.000) & 1.000 & (1.000 , 1.000) \\
  \end{longtable}
\end{center}

\end{appendices}

\end{document}